\documentclass[aps,prd,twocolumn,superscriptaddress]{
revtex4-1}
\usepackage{epsfig}
\usepackage{graphicx}

\usepackage{verbatim} 

\global\long\def\m#1{\mbox{$#1$}}

\begin{document}


\title{Magnetic field amplification by the small-scale
dynamo in the early Universe}


\author{Jacques M. Wagstaff}\email{jwagstaff@hs.uni-hamburg.de}
\author{Robi Banerjee}
\affiliation{Hamburger Sternwarte, Gojenbergsweg 112, 21029 Hamburg, Germany}

\author{Dominik Schleicher}
\affiliation{Institut f\"ur Astrophysik, Georg-August-Universit\"at G\"ottingen, Friedrich-Hund-Platz 1, 37077 G\"ottingen, Germany}

\author{G\"{u}nter Sigl}
\affiliation{II Institut f\"ur Theoretische Physik, Universit\"at Hamburg, Luruper Chaussee 149, 22761 Hamburg, Germany}



\date{\today}

\begin{abstract}
In this paper we show that the Universe is already
strongly magnetized at very early epochs during cosmic
evolution. Our calculations are based on the efficient
amplification of weak magnetic seed fields, which are
unavoidably present in the early Universe, by the turbulent
small-scale dynamo. We identify two mechanisms for the
generation of turbulence in the radiation dominated epoch
where velocity fluctuations are produced by the primordial
density perturbation and by possible first-order phase
transitions at the electroweak or QCD scales.
We show that all the necessities for the small-scale dynamo
to work are fulfilled. Hence, this mechanism, operating 
due to primordial density perturbations, 
guarantees fields with comoving field strength 
\m{B_0\sim10^{-6}\varepsilon^{1/2}}~nG on scales up to
\m{\lambda_c\sim0.1}~pc, where $\varepsilon$ is the saturation efficiency.
The amplification of magnetic seed fields could be even larger if there are first-order 
phase transitions in the early Universe. Where, on scales up to
\m{\lambda_c\sim100}~pc, the comoving field strength due to this mechanism will be
\m{B_0\sim10^{-3}\varepsilon^{1/2}}~nG at the present time. Such
fields, albeit on small scales, can play an important role in structure formation
and could provide an explanation to the apparently
observed magnetic fields in the voids of the large-scale
structure.
\end{abstract}


\maketitle

\section{Introduction}

Magnetic fields of strengths of order a few $\mu$G have been
observed in galaxies at
high and low redshifts, in galaxy clusters and in
superclusters~\cite{BFields_galax,BFields_Bernet,
BFields-Clusters,
BFields_SClusters,greenpeas}. There
is also evidence for strong extragalactic magnetic
fields coming from $\gamma$-ray observations. These studies
place a lower bound on
intergalactic magnetic fields at \m{3\times10^{-7}}~nG
\cite{Neronov:1900zz}, although plasma effects may
complicate the propagation of electromagnetic
cascades~\cite{Broderick_etal_2012,Schlickeiser_etal_2012}.
%

Theoretically, magnetic fields are very likely to have been
generated at
some level in the early Universe through a variety of
mechanisms. On the largest scales, magnetic fields can be
generated during inflation giving today
\m{B_0\sim(10^{-25}-10^{-1})}~nG on a scale of
$1$~Mpc~\cite{Turner:PMF}. However, such mechanisms require
some
modification to the Maxwell theory in order to break its
conformal invariance.
Without such modifications, fields of
present strengths \m{B_0\sim(10^{-20}-10^{-11})}~nG 
could have been generated at the
electroweak (EW) and QCD phase transitions
respectively~\cite{B-phaseT}. In this case, the magnetic
field coherence
length is limited by the particle horizon size at the time
of generation, typically much smaller than $1$~Mpc. At later
times, magnetic fields could have
been generated
through the generation of vorticity in the primordial plasma
\cite{Dolgov:2001nv,Matarrese-B,Takahashi:2005nd,
Kobayashi:2007wd,Fenu:2010kh} (originally proposed by
Harrison~\cite{Harrison}). This mechanism is very natural,
since vorticity in the plasma is unavoidably generated in
the late radiation era through the nonlinear couplings of
first-order density perturbations. The seed fields generated
here are of order \m{B_0\sim10^{-20}}~nG.

%
%

In many cases, the
observed magnetic fields are much stronger than fields
predicted by theories. Therefore, in order to explain
observations, some amplification of the generated seed
fields must have occurred at some
point in the history of the Universe. A popular mechanism
for such amplification is known as the dynamo mechanism.

The dynamo mechanism comes in two broad classes (see
Ref.~\cite{Brandenburg:2004jv} for a review). 
The large-scale dynamo
converts kinetic energy on large
scales into magnetic energy. This mechanism can act if the
conducting fluid flow is highly
helical, inhomogeneous or anisotropic, the typical
example being the differential
rotation of galaxies. This galactic dynamo, which operates
only for spiral galaxies, requires seed fields of order 
\m{B_0\sim10^{-21}}~nG to obtain $\mu$G
strengths today~\cite{Davis:1999bt}. However, this type of
dynamo cannot explain strong fields in much younger
galaxies, in galaxy clusters and superclusters or indeed in
the voids of the large-scale structure.

The second class of dynamo works with stationary,
homogeneous and isotropic turbulence. This mechanism, known
as the small-scale dynamo (SSD), also converts kinetic
energy from turbulent motions into magnetic energy and
typically operates on much smaller scales. 
Magnetic field lines, which are frozen into the
conducting plasma, are stretched, twisted and folded by the
random motions of fluid elements leading to exponential
field amplification. The SSD mechanism has
been applied to the formation of the first stars
and galaxies in the matter dominated Universe
\cite{Schleicher:2010ph,Schober:2013aoa} (see
Ref.~\cite{BFields_Beck} for an early discussion on this
subject), where the turbulent motions arise from
gravitational collapse, accretion and supernovae explosions
(e.g. Refs.~\cite{Balsara2001,Sur2012,Schober:2013aoa,Beck2013,Seifried2014}).
If turbulence is predominantly injected by supernova explosions, 
this mechanism may further explain the observed correlation 
between the star formation rate and the magnetic field strength 
in spiral galaxies \cite{Schleicher2013}.
This mechanism can be highly effective at magnetizing
structures in the early Universe. However, a problem evades
explanation; the large field strengths apparently observed
in the voids of the large-scale
structure~\cite{Neronov:1900zz}.

In this paper we investigate the SSD
amplification of magnetic seed fields in the radiation
dominated Universe. If significant turbulence is generated
in this early epoch, then small magnetic seed
fields could be
amplified very efficiently by the mechanism. Unlike velocity
perturbations, magnetic fields survive through the viscous
damping and free-streaming regimes. Therefore, the SSD
could be an effective mechanism to strongly magnetize the
early Universe prior to structure formation, leading to
strong
intergalactic magnetic fields. We demonstrate that the
conditions are right for efficient dynamo amplification
leading to large
magnetic fields, albeit on very small scales, which could
explain observations and have
an impact on early structure formation.


The structure of our paper is as follows.
In Sec.~\ref{SSD} we give a brief review of the small-scale
dynamo mechanism and the conditions necessary for its
action. In Sec.~\ref{Vel_pert}
we look at two mechanisms for the generation of turbulence
in the radiation dominated era. In Sec.~\ref{sec:evol} we
investigate the amplification of magnetic fields due to SSD
action and look at their subsequent evolution to the
present time. We summarize in Sec.~\ref{sec:sum} and
conclude in Sec.~\ref{sec:conc}.

\section{The small-scale dynamo mechanism}\label{SSD}

The small-scale dynamo (SSD) mechanism is a very efficient
mechanism at converting kinetic energy from
turbulent motions to magnetic
energy~\cite{Batchelor,Kazantsev}. To describe the
mechanism, we
first briefly review the conditions necessary for
turbulence to arise. 

%
%

The kinetic Reynolds number $R_e$ characterizes the
relative importance of the fluid advective and dissipative
terms in the Euler equation. For random motions
correlated on some physical scale $l$ with 
\emph{root-mean-square} (rms) velocity
\m{v_l^{\text{rms}}}, the local kinetic Reynolds number is
given by~\cite{Book:Kolb_Turner,Banerjee:2004df}
\begin{equation}\label{Re}
R_e(l) =
 \left\{
\begin{array}{lr}
&\mbox{$\Large{\frac{v_l^{\text{rms}}l}{\eta_{s}}}$} 
\qquad\text{if}\quad l\gg l_{\text{mfp}}\vspace{5pt}\\
&\mbox{$\Large{\frac{v_l^{\text{rms}}}{\alpha_{\text{d}}l
} } $ }
\qquad\text{if}\quad l\ll l_{\text{mfp}}
\end{array}
 \right.
\end{equation} 
for dissipation due to diffusing particles and
free-streaming particles respectively, where
$l_{\text{mfp}}$ is the particle \emph{mean-free-path}
(mfp). Here, $\eta_{s}$ is the shear viscosity and
$\alpha_{\text{d}}$ is a drag
coefficient due to the occasional
scattering of fluid
particles~\cite{Jedamzik:1996wp,Banerjee:2004df}.
%
%
%
On a given scale $l$, a \textit{viscous} regime corresponds
to \m{R_e(l)\ll1}. Whereas for a turbulent regime
\m{R_e(l)\gg1}, in this case the dissipative
time scale is much greater than the \textit{eddy-turnover}
time scale $\tau_{l}$, where
\begin{equation}
 \tau_{l}\equiv l/v_l^{\text{rms}}=al_c/v_l^{\text{rms}}\,,
\end{equation}
and $l_c$ is a comoving length. 
%
%
%

With the injection of kinetic energy, a turbulent flow
develops almost inevitably when the
kinetic Reynolds numbers are large
enough~\cite{Book:Landau_Lifschitz,Book:Frisch}.
Indeed, there is a critical value for which turbulence is
expected, i.e. \m{R_e(l)\gtrsim R_e^{\text{cr}}\sim10^3}. 
The fundamental reasons for the transition to a turbulent
flow
are not completely understood. However, flow instabilities
always arise when the Reynolds numbers are
larger than the critical value~\cite{Book:Landau_Lifschitz}.
One possible mechanism
for the
triggering of the flow instability is due to thermal
fluctuations~\cite{Tsuge:1974}, which are important in the
radiation epoch, but other mechanisms may exist too.
Indeed, in this paper we will look at turbulence
driven by bubble collisions in first-order phase
transitions.
%
%
%

Turbulence is characterized by a \textit{direct
cascade} of energy from
large scales to small scales. The eddy-turnover
time gives the time over which eddy flows break down to
smaller scales in this direct cascade. With a
continuous injection of kinetic energy (or forcing/stirring
of the
fluid) at the \textit{forcing/stirring}
scale $L$, turbulence
becomes fully developed (or \emph{stationary}) on a time
scale of order the
eddy-turnover time scale at the forcing scale $\tau_L$. In
an expanding Universe, this time scale must be at most given
by the Hubble time i.e. \m{\tau_L=1/H}. Thus, the largest
possible forcing scale is \m{L=v_L^{\text{rms}}/H}, where
\m{v_L^{\text{rms}}} is the typical velocity fluctuation on
the forcing scale $L$. The velocity spectrum for
fully developed turbulence is then given by
\begin{equation}\label{vel_turb_spec}
 v_l^{\text{rms}}=v_L^{\text{rms}}\left(\frac{l}{L}
\right)^\vartheta\,.
\end{equation}
The scaling index $\vartheta$ varies between the two
extremes \m{1/3\leq\vartheta\leq1/2}, where
for
incompressible Kolmogorov type turbulence
\m{\vartheta=1/3} and for highly compressible Burgers
type turbulence \m{\vartheta=1/2}. This spectrum is valid
only on the \textit{inertial range}:
\m{l_{\text{diss}}<l<L}, which is determined by the
turbulent kinetic energy cascade. The dissipative scale
$l_{\text{diss}}$
is the scale at which turbulent velocities are diffused due
to viscosity at the
same rate as they are replenished from larger scales. At
this scale the direct cascade ends. The dissipative scale
can be defined through \m{R_e(l_{\text{diss}})\sim1}, hence
\m{l_{\text{diss}}\sim LR_e(L)^{-1/(\vartheta+1)}} assuming
\m{l_{\text{diss}}\gg l_{\text{mfp}}} in Eq.~(\ref{Re}).
%
%

%
%
%

The SSD mechanism converts this turbulent kinetic energy
to magnetic energy~\cite{Batchelor,Kazantsev}. The
effectiveness of the mechanism depends strongly on three
important environmental factors (i) the Reynolds number;
stronger turbulence is more effective (ii) the turbulent
velocity modes; rotational modes are much more efficient
than longitudinal modes \cite{Federrath:2011qz,SSD-Re_Pm}
and (iii) the Prandtl number \m{P_m\equiv
R_m/R_e=4\pi\sigma\eta_s}, where the Prandtl number is a
measure of the relative importance of the magnetic and
kinetic diffusion. Here, the $R_m$ is the
magnetic Reynolds number given by
\begin{equation}\label{Rm}
 R_m(l)=4\pi \sigma al_cv_l^{\text{rms}}\,,
\end{equation}
and $\sigma$ is the plasma conductivity.

There are two competing effects in the turbulent dynamo
mechanism; magnetic field line stretching and resistive
reconnection. The critical magnetic Reynolds number
\m{R_m^{\text{cr}}} defines the balance between the
stretching and reconnection. For \m{R_m(L)< R_m^{\text{cr}}}
reconnection wins and there is no dynamo and for \m{R_m(L)>
R_m^{\text{cr}}} the stretching wins and the dynamo takes
effect amplifying the magnetic field. Independent of the
Prandtl number, the critical magnetic
Reynolds number is \m{R_m^{\text{cr}}\approx60}
and \m{R_m^{\text{cr}}\approx2700} for Kolmogorov and
Burgers type turbulence
respectively
\cite{Brandenburg:2004jv,SSD-Re_Pm,Schober2012c}. When the
SSD takes effect, the fluctuating component of the magnetic
field grows exponentially
\m{B_{\text{rms}}\propto\exp(\Gamma t)} due to turbulence in
a weakly magnetized plasma. Depending on the Prandtl
number, the growth rate $\Gamma$ scales with either the
kinetic Reynolds number $R_e$ or the magnetic Reynolds
number $R_m$, in particular 
\m{\Gamma\propto R_e^{(1-\vartheta)/(1+\vartheta)}}
and
\m{\Gamma\propto R_m^{(1-\vartheta)/(1+\vartheta)}}
for \m{P_m\gg1} and \m{P_m\ll1}
respectively~\cite{SSD-Re_Pm,Schober2012c}.

There is
a large number of numerical studies in the literature that
have demonstrated the SSD action unambiguously for a
number of
settings~\cite{Federrathnum,Latif:2012aq,Banerjee:2012ch,
Federrath:2011qz,Haugen:2003xp}.
Analytically, the Kazantsev
model (following the formalism by
Refs.~\cite{Subramanian:1997pd,Brandenburg:2004jv})
was developed in order to study the
evolution of magnetic fields in a conducting plasma
containing turbulent motions~\cite{Kazantsev}. The
Kazantsev model can be used to calculate the magnetic field
growth rate and the critical magnetic Reynolds number
required for SSD action. The model considers random
turbulent motions correlated on a scale $l$ with velocity
\m{v_l^{\text{rms}}} and spectrum given in
Eq.~(\ref{vel_turb_spec}) valid on the inertial range
\m{l_{\text{diss}}<l<L}. The spectrum of velocity
fluctuations is assumed to be Gaussian, homogeneous and
isotropic in space and instantaneously correlated in time.
The kinetic Reynolds number must also be larger
than some critical value for which turbulence is expected,
i.e. \m{R_e(L)\gtrsim R_e^{\text{cr}}\sim10^3} 
(this is a conservative estimate for $R_e^{\text{cr}}$,
indeed the SSD action has been observed in cases
where
\m{R_e(L)\lesssim100}~\cite{Banerjee:2012ch,Federrathnum}).
By modeling the turbulent velocity spectrum so that it
behaves as Kolmogorov or Burgers turbulence for scaling
index \m{\vartheta=1/3,1/2} respectively, it can be
shown that the magnetic field growth rate is given by
\cite{SSD-Re_Pm,Bovino-etal:2013}
\begin{equation}\label{gamma}
 \Gamma=\frac{(163-304\vartheta)}{60}
R_e(L)^{(1-\vartheta)/(1+\vartheta)}\tau_L^{-1}\,.
\end{equation}
This result is valid in the large Prandtl number limit
\m{P_m\gg1}, which is relevant to cosmological plasmas.
Since the Reynolds numbers are typically very large, the
magnetic fields can be amplified very rapidly. We also note
that this analytical result for the growth rate
has been verified by Ref.~\cite{Bovino-etal:2013} via a
numerical integration of the Kazantsev equation.

The phase of rapid exponential amplification comes to an end
when the magnetic energy becomes comparable to the kinetic
energy on the dissipative scale \m{l_{\text{diss}}}. The
system then enters a stage of nonlinear growth, where the
magnetic field grows as some power law in time
\m{B_{\text{rms}}(t)\propto
t^{\vartheta/(1-\vartheta)}}~\cite{Schleicher:2013iz}. This
phase lasts until the magnetic field is saturated on the
forcing scale $L$. 
%
%
%
%
Saturation is given by the approximate equipartition
between magnetic and kinetic energy
\m{E_{\text{M}}/E_{\text{kin}}\approx\varepsilon}. This
occurs when
\begin{equation}\label{saturation}
 \langle \mbox{\boldmath
$B$}^2(\mbox{\boldmath
$x$})\rangle\approx4\pi \varepsilon(\rho+p)\langle
\mbox{\boldmath
$v$}^2(\mbox{\boldmath
$x$})\rangle\,,
\end{equation}
where the parameter $\varepsilon$ quantifies the
saturation efficiency. Numerical studies
(for \m{P_m\approx2}) indicate that the SSD
mechanism is more efficient for rotational modes, where the
saturation efficiency $\varepsilon$ is close to
unity~\cite{Federrath:2011qz}. Whereas the saturation level
is lower for compressive modes
\m{\varepsilon\sim10^{-3}-10^{-4}}~\cite{Federrath:2011qz}.
However, further numerical work is required to establish the
saturation level for large Prandtl numbers. So far there
are no analytical results to determine the efficiency
parameter $\varepsilon$.

Here, we stress that the SSD mechanism is a rather generic
phenomenon, in the sense that the mechanism works
independently of the type
of
turbulence~\cite{Gruzinov:1996gm,Schekochihin:2001pj,
SSD-Re_Pm}. In particular, it is interesting to note that
even purely irrotational turbulence can still drive a
small-scale dynamo. This was originally shown by
Ref.~\cite{Schekochihin:2001pj} and later
Ref.~\cite{SSD-Re_Pm} reached similar conclusions. Hence,
the efficient amplification of magnetic fields seems
unavoidable if any kind of turbulence is generated in a
magnetized plasma.

\section{Turbulence in the early
Universe}\label{Vel_pert}

In this section we use Eq.~(\ref{Re}) to calculate the
Reynolds numbers in the radiation dominated (RD) era in
order to identify epochs of turbulence. With large Reynolds
numbers, as we have argued in the previous section that any
injection of
kinetic energy into the plasma
%
%
will lead to a state of fully developed turbulence
for a range of scales. In the RD era, the
kinetic
Reynolds number in the diffusive regime is given by
\begin{equation}\label{Re2}
 R_e(l,T)
=\frac{5g_*(T)}{g_{\nu,\gamma}}
\frac{v_l^{\text{rms}}
l_c}{l_{\text{mfp},c}^{\nu,\gamma}(T)}\,,
\end{equation}
where the shear viscosity
\m{\eta_{s}=(g_{\nu,\gamma}/5g_*)l_{\text{mfp}}^{\nu,\gamma}
} is
determined by the particles of longest mean free
path
$l_{\text{mfp}}^{\nu,\gamma}$, which are either neutrinos
or photons depending on the time. Here, $g_*$ and
$g_{\nu,\gamma}$ are the total and component number
of effective relativistic degrees of freedom.

In the very early Universe, before neutrino decoupling
\m{T\gtrsim2.6}~MeV, neutrinos
have the longest mfp and are thus most efficient at
transporting momentum and heat. At high temperatures the
shear viscosity due to neutrinos is low and the plasma could
be in a turbulent regime \m{R_e\gg1}. At this time, the
comoving mfp is~\cite{Jedamzik:1996wp}
 \begin{equation}\label{l_c-neutrinos}
 l_{\text{mfp},c}^\nu\simeq
\frac{a^{-1}}{G^2_FT^2(n_{l}+n_q)}\,,
\end{equation}
which is proportional to $1/T^4$. Here
\m{n_l=6g_l\zeta(3)T^3/7\pi^2}
and \m{n_q=6g_q\zeta(3)T^3/7\pi^2} are the lepton
and quark number densities, $g_{l,q}$ are the number of
degrees of freedom for relativistic leptons and quarks,
$\zeta$ is the Riemann zeta function and
$G_F$ is the Fermi constant.
However, the neutrino mfp increases as the Universe
expands and cools,
leading to a viscous regime \m{R_e<1} (see e.g.
Ref.~\cite{Banerjee:2004df}). Eventually the neutrinos
decouple at \m{T_{\text{dec}}\simeq2.6}~MeV. From here on,
momentum
and heat is effectively transported by the photons. At early
times, photons generate a small shear viscosity in the
plasma and the fluid flow could become turbulent once
again. The comoving photon mfp is given
by~\cite{Jedamzik:1994dd}
%
%
\begin{equation}\label{l_c-photons}
 l_{\text{mfp},c}^\gamma\simeq
\frac{a^{-1}}{\sigma_T(n^2_{\textrm{\tiny
pair}}+n^2_e)^{1/2}}\,,
\end{equation}
where \m{\sigma_T=8\pi\alpha^2/3m_e^2} is the
Thomson cross section, \m{\alpha\approx1/137} is
the fine structure constant and $m_e$ is the electron mass.
During this epoch, the number
densities 
$n_{\textrm{\tiny pair}}$ and $n_e$ of $e^{\pm}$ pairs and
free electrons respectively are given
by~\cite{Jedamzik:1994dd}
%
%
%
\begin{eqnarray}
 n_{\textrm{\tiny pair}}&\approx&
\left(\frac{2m_e
T}{\pi}\right)^{3/2}\exp\left(-\frac{m_e}{T}
\right)
\left(1+\frac{15}{8}\frac{T}{m_e}\right)\,,\label{n_pairs}\\
n_e&=&X_e\frac{\Omega_b\rho_0}{m_{p}}\left(\frac{T}{T_0}
\right)^ { 3 }\,,\label{n_free}
\end{eqnarray}
where
$m_{p}$ is the proton mass, the baryon fraction and present
day density product is
\m{\Omega_b\rho_0\simeq1.81\times10^{-12}~\text{eV}^4}
\cite{Planck_param}, \m{T_0\simeq2.725}~K is the present day
photon temperature and 
the ionization fraction is \m{X_e=1} in the RD era.

Once the
temperature decreases below the electron mass
\m{T<m_e\simeq0.511}~MeV, $e^\pm$ pairs begin to
annihilate and the photon mfp increases rapidly.
The $e^\pm$ annihilation completes at
around \m{T\simeq20}~keV.
After this, as the temperature drops further, photons begin
to diffuse followed by photon drag and the
fluid is in a viscous regime once
again~\cite{Jedamzik:1996wp,Banerjee:2004df}. Hence, there
are two
epochs in the RD era, before and after
neutrino decoupling, where the Reynolds numbers could be
large and turbulence is potentially fully developed for a
range of scales. 

Diffusing particles also damp away velocity
fluctuations (see Sec.~\ref{Damping}). Therefore,
before diffusion sets in, there is always a possibility for
the plasma to be in a turbulent state if the considered
scales are large enough. However, the
eddy-turnover time increases for larger scales. We must
therefore look at the evolution of all relevant scales
carefully in order to establish whether or not turbulence is
possible. To complete the calculation for the
Reynolds numbers, we must estimate the turbulent velocity
fluctuations in the early Universe. In the next two
subsections we present two mechanisms for the generation of
turbulence in the RD era.

\subsection{Turbulence from primordial density
perturbations}\label{sec:turb_pdp}

To explain the formation of the large-scale
structure in the Universe observed today, a primordial
density perturbation of magnitude
\m{\delta\rho/\rho\sim10^{-5}} is
required at the time of matter-radiation
equality~\cite{Book:Kolb_Turner}. The
primordial density perturbation is thought to have been
generated at a much earlier time and therefore must be
present in the very
early Universe during the RD era.
Cosmic inflation provides the most compelling explanation
for the origin of the
primordial perturbation~\cite{Book:Lyth2009}.

Well before horizon entry, the primordial curvature
perturbation, which determines the gravitational potential
$\Phi$, remains constant and given by the initial
condition $\Phi_0$. In the Newtonian gauge, its equation of
motion is 
\m{\Phi''+3\mathcal{H}(1+w)\Phi'-w\nabla^2\Phi=0},
where \m{p=w\rho}, \m{\mathcal{H}=a'/a} and
\m{'\equiv\partial/\partial\eta} 
with conformal time \m{\eta}~\cite{sasaki-cpt}. In the RD
era
\m{w=1/3}, the
solution for the Fourier modes
reads \m{\Phi(\mbox{\boldmath $k$},\eta)
 =3[j_1(y)/y]\Phi_0(\mbox{\boldmath
$k$})}, where \m{y\equiv k\eta/\sqrt{3}},
$k$ is the comoving wave number and
\m{j_1(y)=\sin y/y^2-\cos y/y} is the first
spherical Bessel function. As a Fourier mode of the
gravitational potential reenters the horizon during the RD
era, it begins to oscillate with an amplitude decreasing as
\m{1/y^2\propto1/t}.

The initial conditions, which are probed by observations,
are given by the two-point correlation function
\begin{equation}
 \langle\Phi_0(\mbox{\boldmath
$k$}_1)\Phi^*_0(\mbox{\boldmath
$k$}_2)\rangle=
 (2\pi)^3 P_\Phi(k)\delta^3(\mbox{\boldmath
$k$}_1-\mbox{\boldmath $k$}_2),
\end{equation}
where
\m{P_\Phi(k)=(2\pi^2/k^3)(9/25)
\Delta^2_{\mathcal{R}}(k_0)
 \left(k/k_0\right)^{n_s-1}}.
The results of the Planck mission give 
\m{\Delta^2_{\mathcal{R}}(k_0)=2.215\times10^{-9}} and
\m{n_s\simeq0.96} for the pivot scale
\m{k_0=0.05~\textrm{Mpc}^{-1}}~\cite{Planck_param}.

Perturbations in the fluid 3-velocity field are generated by
the density perturbations. At first-order in density
perturbations, the fluid
velocity perturbation is purely irrotational (curl free)
with Fourier modes~\cite{sasaki-cpt}
\begin{eqnarray}
 v_i(\mbox{\boldmath $k$},\eta)&=&
-\frac{ik_i}{2\mathcal{H}^2}
\left[\Phi'(\mbox{\boldmath
$k$},\eta)
+\mathcal{H}\Phi(\mbox{\boldmath $k$},\eta)\right]\\
&=&
 -i\frac{3\sqrt{3}}{2}\hat k_i
 \left[\sin y -2j_1(y)\right]\Phi_0(k)\,. \label{v_i-k}
\end{eqnarray}
These modes $v_i$ oscillate with the
density perturbation, but have a term which does
not decay with the expansion.

\begin{figure}[h!]
\includegraphics[width=80mm,trim=20 20
20 20]{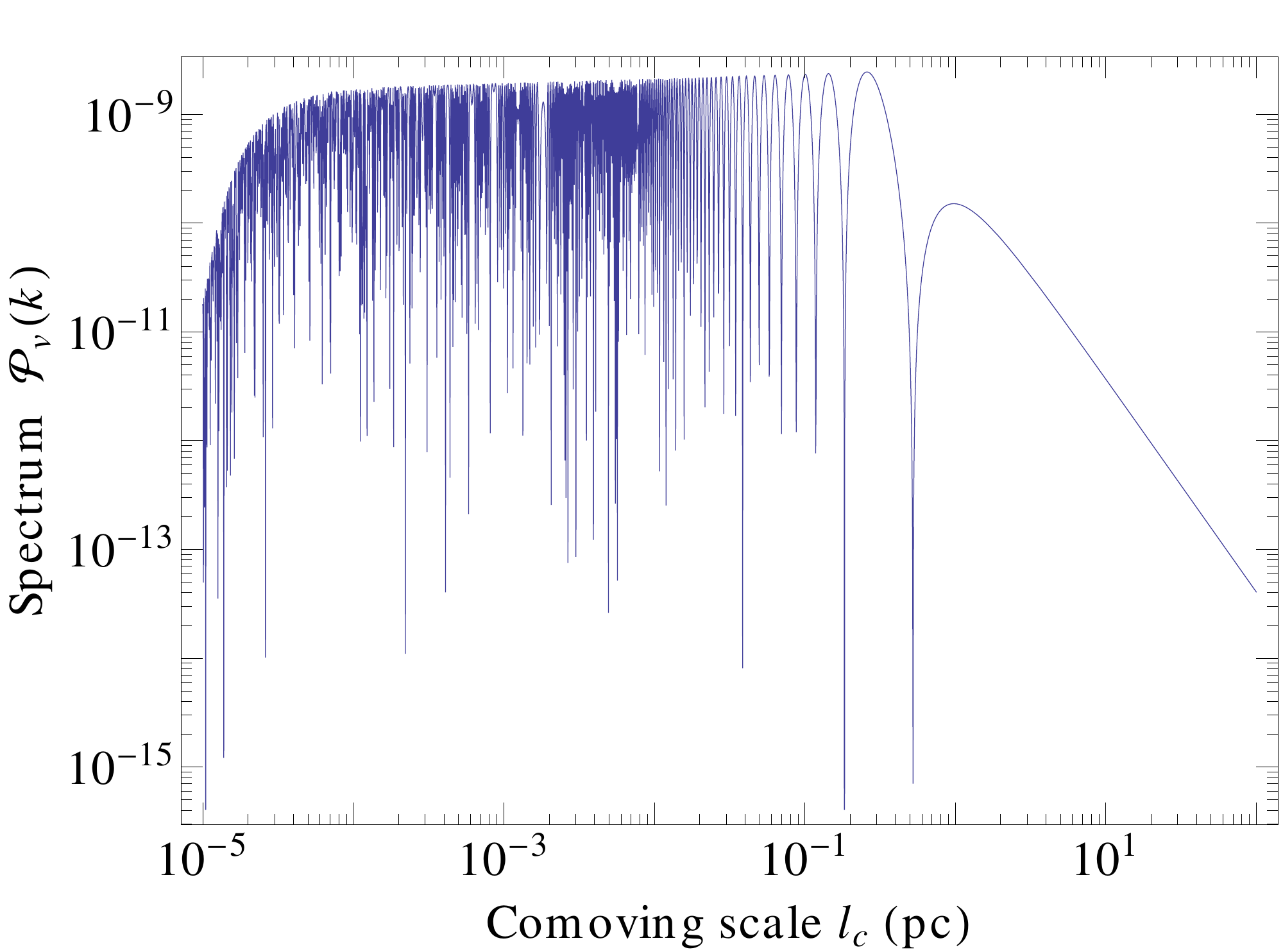}
\caption{\label{fig1} The spectrum of velocity
perturbations, given in Eq.~(\ref{P_v}), at a temperature
\m{T\simeq0.21}~GeV. The figure shows how the velocity
perturbations are generated upon horizon entry (at around
\m{l_c\simeq0.31}pc) and are scale invariant until
damping due to neutrino diffusion takes over at around
\m{l_c\simeq1.5\times10^{-5}}pc (see Sec.~\ref{Damping}).
}
\end{figure}

Let us define the spectrum of
velocity perturbations in Fourier space by the two-point
correlation function
\begin{equation}\label{v2point}
 \langle v_i(\mbox{\boldmath $k$}_1,
\eta)v^*_i(\mbox{\boldmath $k$}_2, \eta)\rangle=
 (2\pi)^3 \frac{2\pi^2}{k^3}\mathcal{P}_v(k)
 \delta^3(\mbox{\boldmath
$k$}_1-\mbox{\boldmath $k$}_2)\,.
\end{equation}
Hence, with Eq.~(\ref{v_i-k}) and the spectrum $P_\Phi$ we
find 
\begin{equation}\label{P_v}
 \mathcal{P}_v(k)=
\frac{243}{100}\left[\sin y -2j_1(y)\right]^2
 \Delta^2_{\mathcal{R}}(k_0)
 \left(k/k_0\right)^{n_s-1}\,.
\end{equation}
%
%
%
The velocity spectrum oscillates rapidly for subhorizon
scales \m{y,k/\mathcal{H}\gg1}, therefore we can
average $\mathcal{P}_v(k)$ over many oscillations. For a
scale invariant primordial spectrum \m{n_s=1} we find
\m{\overline{\mathcal{P}}_v(k)\simeq\frac{243}{200
}\Delta^2_{\mathcal{R}}(k_0)}. Hence, on subhorizon scales,
the spectrum of
velocity perturbations generated by first-order
density perturbations is
isotropic, homogeneous, Gaussian
and to a good approximation scale invariant, see
Fig.~\ref{fig1}.

The \emph{root-mean-square} (rms) velocity is then given by
\begin{equation}
 v^{\text{rms}}\equiv
\sqrt{\langle \mbox{\boldmath $v$}^2(\mbox{\boldmath
$x$})\rangle}=\left(\int_0^\infty \mathcal
P_v(k)\frac{dk}{k}\right)^{1/2}\,.
\end{equation}
It will also be useful to define
%
%
the rms
velocity on a given comoving length scale \m{l_c=2\pi/k_l}
by
\begin{equation}
 v_l^{\text{rms}}\simeq\left(\int_{k_l}^{k_{\text{diss}}}
\mathcal
P_v(k)\frac{dk}{k}\right)^{1/2}\,,
\end{equation}
where \m{l_{\text{diss},c}=2\pi/k_{\text{diss}}} is a
cut-off or dissipative scale. The typical value of the
velocity
perturbation
is therefore
\begin{equation}\label{v_rms}
 v_l^{\text{rms}}\sim\sqrt{\overline{\mathcal{P}}_v(k)}
\simeq
\sqrt{\Delta^2_{\mathcal{R}}(k_0)}
 \simeq 5\times10^{-5},
\end{equation}
where natural units are used such that \m{c=1}. 
%
%
%

In the radiation dominated era, new $k$-modes of the
primordial density
perturbation, which generate velocity perturbations, are
continuously reentering the horizon. This continuous
production of velocity perturbations can be seen as the
continuous \textit{forcing} of the fluid on the largest
scales. Thereby, if the Reynolds numbers are large enough,
turbulent flow will occur. The velocity perturbations
forcing the fluid are purely longitudinal in this case.
However, rotational modes will be generated at second order
in cosmological perturbations~\cite{Lewis:2004kg,
Lu:2007cj, Lu:2008ju, Christopherson:2010dw, Ichiki:2011ah}.
In any case, with \m{R_e\gg1}, nonlinear
interactions can play a role in generating turbulence with
both longitudinal and rotational fluid
motions. Therefore, a state of fully developed (or
stationary) turbulence can be expected on a time scale of
order the eddy-turnover time scale at the forcing scale
$\tau_L$ i.e. \m{\tau_L=1/H} in an expanding Universe (see
Sec.~\ref{SSD}). Hence, for all scales below
\m{L=v_L^{\text{rms}}/H}, there
are many eddy-turnover times per Hubble time
\m{\tau_l\lesssim 1/H}. This condition ensures interactions
between eddy flows leading to fully developed
turbulence with a spectrum given
in Eq.~(\ref{vel_turb_spec}). A second condition is that the
Reynolds number on the
forcing scale $L$ is larger than some critical value for
which turbulence is expected, i.e.
\m{R_e(L)\gtrsim R_e^{\text{cr}}\sim10^3}.
With these two conditions we find the range of scales,
corresponding to the inertial range,
\m{l_{\text{diss}}\lesssim l\lesssim L}, where
\m{R_e(l_{\text{diss}})\sim1}.
In Sec.~\ref{sec:turbRD} we establish when, in
the RD era, and on what scales, the Reynolds numbers are
large given the turbulent velocity fluctuations generated by
the primordial density perturbations.

\subsection{Turbulence injected from phase
transitions}\label{sec:PT}

In this section we briefly describe another mechanism for
the generation of turbulence in the RD era. The mechanism
occurs during first-order phase
transitions when bubbles of the new phase collide and
merge~\cite{Steinhardt1982,
Kamionkowski:1993fg,Kosowsky:2001xp}. 
In the early Universe, the electroweak and QCD phase transitions 
are potentially first order, although under early Universe 
conditions with very small chemical potentials the QCD transition 
is a smooth transition whereas the electroweak transition could 
be first order in certain Standard Model extensions.
These violent phenomena can inject large kinetic
energy into the plasma, thereby generating turbulence and
allowing the possibility of small-scale dynamo action. 

The characteristic time scale for
the phase transition is
given by the rate of bubble nucleation $\beta^{-1}$. Here
$\beta$ is expected to be \m{\beta\sim
100H}~\cite{Kosowsky:2001xp}. The
largest bubbles reach a size \m{\beta^{-1}v_b} by
the end of the phase transition, where $v_b$ is the bubble
wall expansion velocity. Thus, we take \m{L\simeq
\beta^{-1}v_b} as the
largest stirring scale and \m{\tau_{\text{stir}}=\beta^{-1}}
as the stirring time scale~\cite{Kosowsky:2001xp}.
The phase boundary can propagate via two modes,
\emph{deflagration} and \emph{detonation}, where the wall
velocity $v_b$ is subsonic and supersonic
respectively~\cite{Steinhardt1982,Kamionkowski:1993fg}. 
It has been argued in the literature that deflagrations are
unstable to becoming detonations via bubble wall
instabilities~\cite{Kamionkowski1992}. Hence, for simplicity
we will only consider detonations, where
the wall velocity is fully
determined and given by~\cite{Steinhardt1982}
\begin{equation}
 v_b(\alpha)=\frac{1}{1+\alpha} 
\left(\frac{1}{\sqrt{3}}+\sqrt{\alpha^2+\frac{2\alpha}{3}}
\right)\,.
\end{equation}
Here
\m{\alpha\equiv\rho_{\text{vac}}/\rho_{\text{thermal}}}
determines the strength of the phase transition. In this
case, the fraction of vacuum energy converted to kinetic
energy \m{\kappa\equiv\rho_{\text{kin}}/\rho_{\text{vac}}}
takes the form~\cite{Kamionkowski:1993fg}
\begin{equation}
 \kappa(\alpha)=\frac{1}{1+0.72\alpha} 
\left(0.72\alpha+\frac{4}{27}\sqrt{\frac{3\alpha}{2}}
\right)\,.
\end{equation}

For phase transitions that give large stirring times
compared to the eddy turnover time of the largest scale
\m{\tau_{\text{stir}}\gg\tau_L}, a direct cascade of energy
is set up and a state of fully developed turbulence is
established in a time scale $\tau_L$ and can be
expected for a duration time
$\tau_{\text{stir}}$~\cite{Kosowsky:2001xp}. 
Since the rate of energy dissipation is equal to the mean
input power in stationary turbulence, the
amplitude of the Kolmogorov spectrum can be easily
determined. This calculation is done in
Ref.~\cite{Kosowsky:2001xp}, where they show that 
\m{v_L^{\text{rms}}\simeq(\alpha\kappa v_b)^{1/3}} and
argue that \m{v_L^{\text{rms}}\lesssim1/\sqrt3}. Hence, the
condition for this simpler case
\m{\tau_{\text{stir}}\gg\tau_L} is translated to
\m{3v_b\ll\sqrt{2\alpha\kappa}}
\cite{Kosowsky:2001xp,Caprini:2006jb} which is satisfied
only for \m{\alpha\gtrsim1}. Therefore, in this case of
strong detonation \m{\alpha\gtrsim1} we have
\m{v_L^{\text{rms}}\sim1}~\cite{Kamionkowski:1993fg}.
However, if \m{\tau_{\text{stir}}\lesssim\tau_L}, a state of
turbulence can still be expected~\cite{Kosowsky:2001xp}. The
stirring corresponds to an impulsive force acting on the
plasma that will cascade down
to smaller scales. Eddy flows on large scales $L$ act as a
source for eddies on smaller scales for a duration time
$\tau_L$. Following Ref.~\cite{Caprini:2006jb}, in the time
scale $\tau_L$, we neglect the decay of the turbulence and
assume a state of fully developed turbulence for
a duration time $\tau_L$~\cite{Kosowsky:2001xp}. Numerical
work in
Ref.~\cite{Hindmarsh:2013xza} has established that kinetic
energy in the form of acoustic
waves persist well beyond the time of the phase
transition. The nonlinear interaction of these acoustic
waves could also be a source of turbulence on larger time
scales. In the weak detonation limit \m{\alpha\lesssim1},
we find~\cite{Kamionkowski:1993fg,Kosowsky:2001xp}
%
%
\begin{equation}
 v_L^{\text{rms}}\simeq
\frac{\sqrt{2\alpha\kappa}}{3(2\pi)^{4/3}}\,.
\end{equation}
Hence, for first-order phase transitions of strengths in the
range \m{\alpha\sim(10^{-5}-10^{-1})}, we
find turbulent velocities 
\begin{equation}\label{turb_vel_PT}
v_L^{\text{rms}}\sim(10^{-4}-10^{-1})\,.
\end{equation}
In this case, the turbulence 
is expected to be of Kolmogorov type.

%

\subsection{Damping of turbulence}\label{Damping}

Velocity perturbations of the baryonic fluid are damped
below a scale $l_D$ due to
particles diffusing out of overdense regions. The damping is
very efficient and given by (see for example
Ref.~\cite{Book:Dodelson})
\begin{equation}\label{damping}
 v\propto
\exp\left[-\left(\frac{l_D^{\nu,\gamma}}{l_c}\right)^2
\right]\,,
\end{equation}
where $l_c$ is a comoving length scale.
%
%
%
In the RD era, the comoving damping scale due to neutrinos
or photons random walking out of perturbations is given
by~\cite{Book:Kolb_Turner,Book:Dodelson}
\begin{equation}\label{l_Damping}
 (l_D^{\nu,\gamma})^2\simeq\int_0^t
\frac{l^{\nu,\gamma}_{\text{mfp},c}(t')}{a(t')}\textrm{d}t'
\,,
\end{equation}
where $l_{\text{mfp},c}^{\nu,\gamma}$ is the comoving
particle \emph{mean-free-path} (mfp). The efficient damping
of velocity perturbations is seen in the spectrum in
Fig.~\ref{fig1}. This important effect must be
considered carefully when we come to investigate the scales
of turbulence in the RD era (see next
section). Here, we briefly note that although turbulent
velocity fluctuations are efficiently damped due to
diffusing particles, magnetic fields become
overdamped and survive through such viscous and
free-streaming regimes (this effect is discussed in detail
in
Sec.~\ref{sec:sub-evo})~\cite{Jedamzik:1996wp,
Banerjee:2004df,Durrer:2013pga}.

%

\subsection{The scales of turbulence}\label{sec:turbRD}

\begin{figure}[h!]
\includegraphics[width=80mm,trim=20 20 20
20]{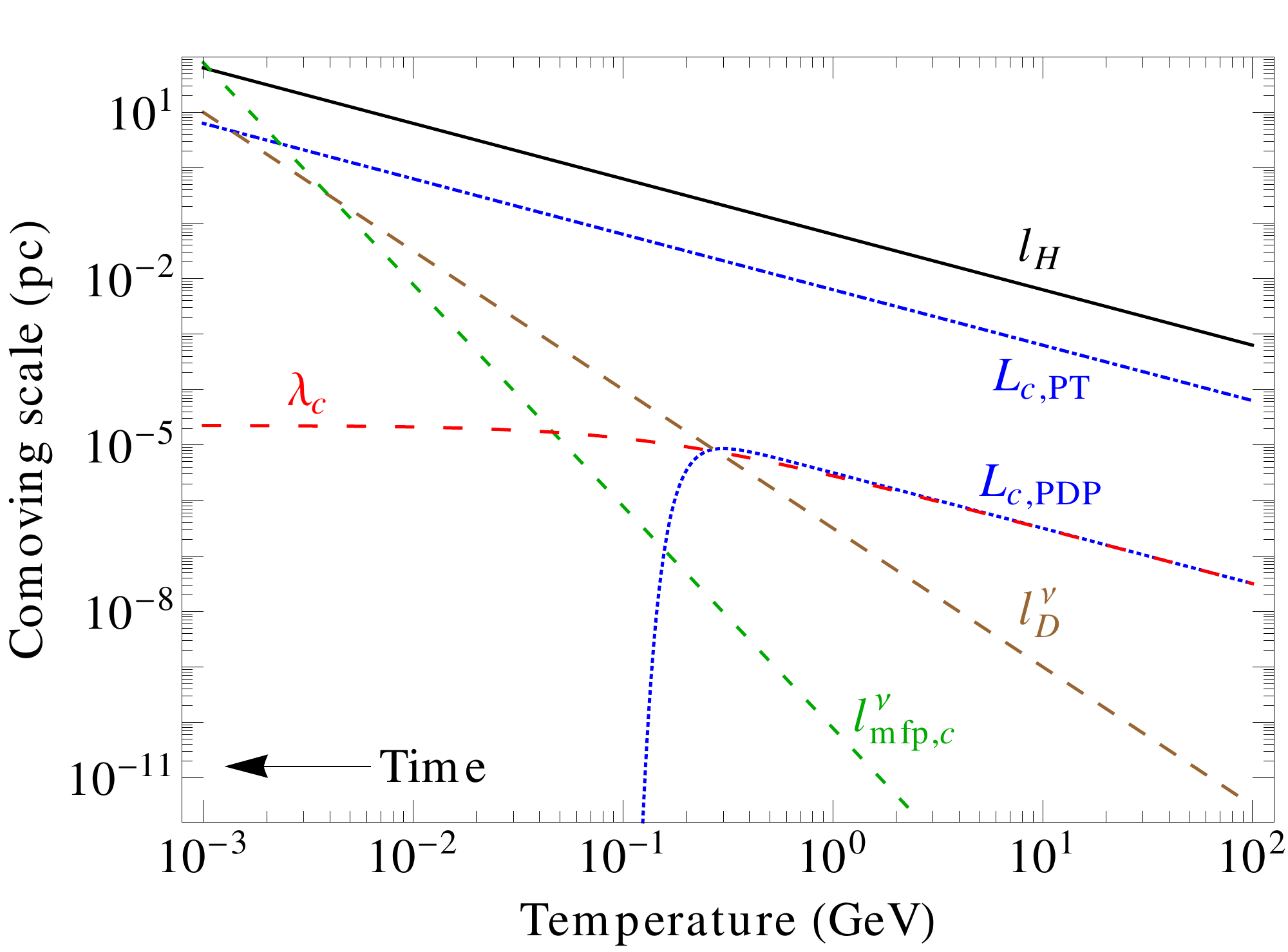}
\caption{\label{fig2} This figure shows the
evolution of relevant comoving scales from the EW scale
\m{T_{\text{EW}}\sim100~}GeV to the time
of neutrino decoupling at \m{T_{\text{dec}}\simeq2.6~}MeV.
In this early epoch the neutrinos generate the plasma
viscosity. The QCD phase transition occurs at around
\m{T_{\text{QCD}}\simeq200}~MeV. Here, \m{l_H=1/aH} is the
Hubble scale (solid, black), \m{l_{\text{mfp},c}^\nu} 
is the neutrino mean-free-path (dashed, green) and $l_D^\nu$ is the damping
scale due to neutrino diffusion 
given by Eq.~(\ref{l_Damping}) (dashed, brown). For
turbulence generated by the primordial density perturbation (PDP)
and first-order phase transitions (PT), the largest stirring
scales \m{L_c=v_L^{\text{rms}}/aH} are shown
with the values
\m{v_L^{\text{rms}}} from eqs.~(\ref{v_rms}) and
(\ref{turb_vel_PT}) (dotted, blue) and (dotdashed, blue)
respectively. In Eq.~(\ref{turb_vel_PT}), the upper value
for \m{v_L^{\text{rms}}} is used. Although the turbulent 
motions from PDP become completely damped below \m{T\simeq0.2}~GeV, the magnetic
field gets \emph{frozen-in} with integral scale $\lambda_c$ (dashed, red).
}
\end{figure}

The evolution of relevant
scales from the time of the electroweak (EW)
scale \m{T_{\text{EW}}\sim100~}GeV to the time
of neutrino decoupling \m{T_{\text{dec}}\simeq2.6~}MeV is
shown in
Fig.~\ref{fig2}. In this
early epoch the neutrinos generate the plasma viscosity.
The QCD phase transition occurs at around
\m{T_{\text{QCD}}\simeq200}~MeV. The scales of
interest are the comoving Hubble scale \m{l_H=1/aH}, the
largest stirring scale \m{L_c=v_L^{\text{rms}}/aH}
with the values \m{v_L^{\text{rms}}} from primordial density
perturbations (PDP)  and first-order phase transitions (PT)
in eqs.~(\ref{v_rms}) and (\ref{turb_vel_PT}) respectively, and the
damping scale due to neutrino diffusion $l_D^\nu$ 
[cf. Eq.~(\ref{l_Damping})]. Here, we assume that the
damping scale due to neutrino diffusion is the only relevant
damping scale at this time i.e. the velocity perturbations
on small scales are not damped
due to physical processes at higher temperatures. 
Indeed, this is a safe assumption since neutrinos are the
most weakly interacting particles in the Standard Model and
at early times are the most efficient heat transporters.

\begin{figure}[h!]
\includegraphics[width=80mm,trim=20 20 20
20]{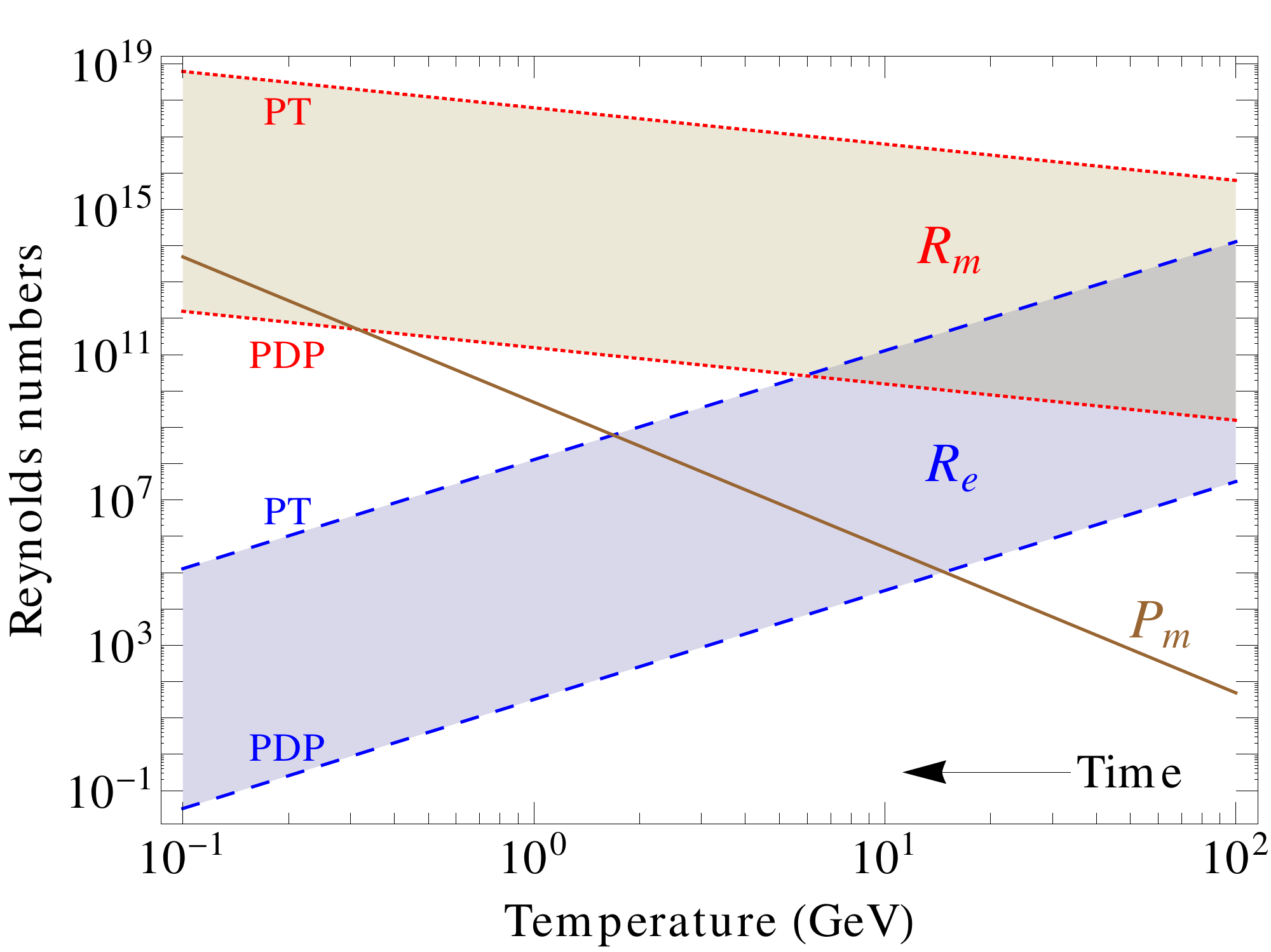}
\caption{\label{fig3} This figure shows the
evolution of the different Reynolds numbers from the
EW scale \m{T_{\text{EW}}\sim100~}GeV to the
QCD scale \m{T_{\text{QCD}}\simeq200}~MeV. The kinetic
Reynolds number $R_e(L_c)$ (dashed, blue) is determined from
eqs.~(\ref{Re2}) and (\ref{l_c-neutrinos})
and the magnetic Reynolds number $R_m(L_c)$
(dotted, red) from eqs.~(\ref{Rm}) and (\ref{conduc_highT}).
The lower lines for the
Reynolds numbers correspond to turbulence generated by the
primordial density perturbations (PDP) i.e. using the (undamped)
value for
\m{v_L^{\text{rms}}} from Eq.~(\ref{v_rms}). The upper lines
for the Reynolds numbers correspond to
turbulence generated by first-order phase transitions (PT) with
the (undamped) value 
\m{v_L^{\text{rms}}\sim10^{-1}} from
Eq.~(\ref{turb_vel_PT}). This figure clearly
shows that \m{R_e\gg1} in all cases, suggesting that the
plasma is in a
state of fully developed turbulence during this time. The
figure also shows the Prandtl numbers
\m{P_m=R_m/R_e\gg1} (solid, brown).}
\end{figure}

Let us first consider the turbulence generated by the
primordial density perturbations (PDP).
From Fig.~\ref{fig2} we can see that for
\m{T\gtrsim0.2}~GeV the stirring scale \m{L_{c,\text{PDP}}} (the lower blue
dotted line in the figure) is
larger than the damping scale $l_D^\nu$. Hence, the velocity
perturbations are not damped and we can use the value
given in Eq.~(\ref{v_rms}). With this value for the
$v_L^{\text{rms}}$ we can calculate the Reynolds
numbers $R_e(L_c)$ from Eq.~(\ref{Re2}), these are shown in
Fig.~\ref{fig3} (the lower dotted blue line). We find that
\m{R_e(L_c)\gg1} for
\m{0.2\lesssim T/\text{GeV}\lesssim100}. The largest
stirring scale, over
which large Reynolds numbers are found, is roughly given at
\m{T\simeq0.2}~GeV i.e.
\m{L_{c,\text{PDP}}\sim10^{-5}}pc. Hence,
at these times, between the damping scale $l_D^\nu$ and
$L_{c,\text{PDP}}$, we expect a state of
fully developed turbulence.
%
However, for \m{T\lesssim0.2}~GeV, the scale $L_{c,\text{PDP}}$ is below
the damping scale $l_D^\nu$, which means that velocity
perturbations generated by the primordial density
perturbation are exponentially damped [cf.
Eq.~(\ref{damping})] and
so are the Reynolds numbers. Thus, below this
temperature, the plasma is in a viscous regime and there is
no turbulence.

For turbulence generated by first-order phase transitions (PT),
the stirring scale can be much larger (the upper dotdashed
blue line in Fig.~\ref{fig2}), since the velocity
fluctuations can be much stronger [cf.
Eq.~(\ref{turb_vel_PT})]. Hence, if the phase transition
occurs at any time in the epoch between
\m{T\sim(10^2-10^{-3})}~GeV, a state of turbulence can be
expected. In Fig.~\ref{fig3}, the Reynolds numbers are shown
to be very large between the EW and QCD scales, indicating a
highly turbulent state. The largest stirring scale, over
which large Reynolds numbers are found, is roughly given by
the horizon size at that time of the phase transition:
\m{1/aH|_{\text{QCD}}\sim0.1}~pc
and \m{1/aH|_{\text{EW}}\sim10^{-4}}~pc for
the QCD and EW phase transitions respectively.

\begin{figure}[h!]
\includegraphics[width=80mm,trim=20 20 20 20]{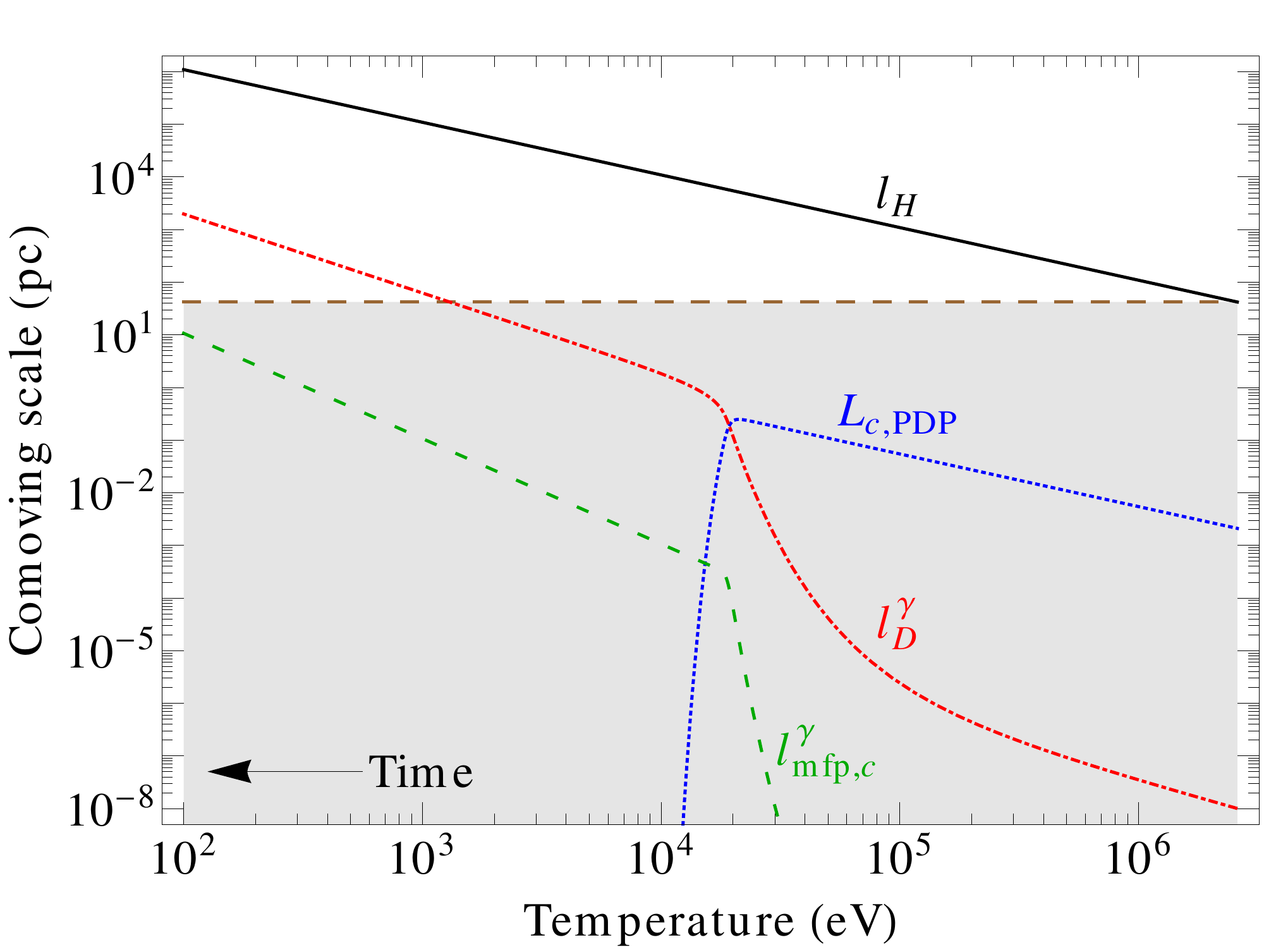}
\caption{\label{fig4} This figure shows the
evolution of relevant comoving scales from the time of
neutrino
decoupling at \m{T_{\text{dec}}\simeq2.6~}MeV to a time long
after
$e^\pm$ annihilation at \m{T\simeq100~}eV. In this epoch
the photons generate the plasma viscosity. The time of
$e^\pm$ annihilation occurs at \m{m_e\gtrsim
T\gtrsim20~}keV. Here, \m{l_H=1/aH} is the
Hubble scale (solid, black), \m{l_{\text{mfp},c}^\gamma} 
is the photon mean-free-path (dashed, green), 
\m{L_{c,\text{PDP}}=v_L^{\text{rms}}/aH}
(dotted, blue) with the value
\m{v_L^{\text{rms}}}
from Eq.~(\ref{v_rms}), and $l_D^\gamma$ is
the damping scale due to photons (dotdashed, red) given by Eq.~(\ref{l_Damping}). 
Velocity fluctuations below the scale \m{l_D^\nu\approx1/aH|_{\text{dec}}\simeq42}~pc 
(shaded area) are damped due to neutrino diffusion at an earlier time, see Fig.~\ref{fig2}.
Hence, below \m{l_c\approx42}~pc,
the velocity perturbations generated by the primordial density perturbations are
completely erased and no turbulence can be generated in this epoch.
}
\end{figure}

The evolution of relevant
scales from the time of neutrino
decoupling \m{T_{\text{dec}}\simeq2.6~}MeV to a time long
after
$e^\pm$ annihilation \m{T\simeq100~}eV
is shown in Fig.~\ref{fig4}. In this
epoch the photons generate the plasma viscosity.
The scales of interest are the comoving Hubble scale
\m{l_H=1/aH}, the
largest stirring scale \m{L_{c,\text{PDP}}=v_L^{\text{rms}}/aH}
with the value \m{v_L^{\text{rms}}} from
Eq.~(\ref{v_rms}) and both damping scales
$l_D^{\nu,\gamma}$ given by Eq.~(\ref{l_Damping}). The
largest damping scale due to neutrino diffusion (which
occurs at an earlier time) is approximately given by the
particle horizon at the time of neutrino decoupling i.e.
\m{l_D^\nu\approx1/aH|_{\text{dec}}\simeq42}~pc 
\cite{Jedamzik:1996wp}. Here, we
assume that, if there is indeed a first-order phase
transition, it would have occurred at a much
earlier time. Therefore, in this epoch, turbulence can only
be generated by the primordial density perturbations.

However, with the value of the rms velocity from
Eq.~(\ref{v_rms}) damped only by photon diffusion, we
can clearly see from Fig.~\ref{fig4} that the scale 
\m{L_{c,\text{PDP}}} is below the 
largest damping scale due to neutrino diffusion
\m{l_D^\nu\approx42}~pc throughout this epoch.
Therefore, the velocity perturbations are efficiently damped
by particle diffusion
[cf. Eq.~(\ref{damping})]. Indeed, on the scale $L_{c,\text{PDP}}$, the
velocity perturbations generated by the primordial density
perturbations are completely erased due to neutrino
diffusion at this time. Hence, the kinetic Reynolds numbers
become vanishingly small even when the shear viscosity is
very small. Thus, at these temperatures, the plasma is in a
viscous regime and there is no turbulence.

\section{Evolution of cosmological magnetic
fields}\label{sec:evol}
In this section we consider the cosmological evolution of
magnetic fields from the time of their generation to the
present day (see for example
Refs.~\cite{Jedamzik:1996wp,Banerjee:2004df,Durrer:2013pga}
). We first consider the amplification of magnetic seed
fields due
to small-scale dynamo (SSD) action in the early Universe.
Then, we consider the
subsequent evolution to the
present time. In order to
compare with observations, it is important to
theoretically determine the final magnetic field strength
and coherence length.

\subsection{Amplification by small-scale dynamo
action}\label{SSDamp}

In Sec.~\ref{Vel_pert} we identified two mechanisms that
generate turbulence in the early Universe. In a turbulent
and weakly magnetized plasma, small magnetic seed fields
can be amplified exponentially through the SSD action (as
described in Sec.~\ref{SSD}). Let us now assume that small
magnetic seed fields exist at the time of the EW
phase transition. These fields may have been generated at
the phase transition \cite{B-phaseT} or at an earlier time
(for example
during inflation~\cite{Turner:PMF}). We now investigate the
possibility of
SSD action in the RD era.

The injection of kinetic energy, together with large
Reynolds numbers, leads to a state of fully developed
turbulence. In the previous section we found that turbulence
is expected due to primordial
density perturbations or first-order
phase transitions at high temperatures, between
\m{0.2\lesssim T/\text{GeV}\lesssim100} (see
Fig.~\ref{fig3}). Indeed, we expect fully developed
turbulence below the stirring scale
\m{L_c=v_L^{\text{rms}}/aH}. At these temperatures, 
the conductivity $\sigma$ is
given by~\cite{Ahonen:1996nq}
\begin{equation}\label{conduc_highT}
 0.76T\lesssim\sigma\lesssim6.7T\,,
\end{equation}
where the larger value corresponds to the upper temperature
bound. With the above we can calculate the Prandtl numbers
and the
magnetic Reynolds numbers. These
are also shown in Fig.~\ref{fig3}. The Prandtl numbers at
these times are
very large, \m{P_m\sim(10^{2}-10^{12})},
which means that we can neglect
dissipative effects due to finite conductivity throughout
the epoch of interest. From Eq.~(\ref{Rm}), we find the
magnetic Reynolds numbers
\m{R_m(L_c)\sim(10^{9}-10^{12})} for turbulence generated
by primordial density perturbations and a maximum range of 
\m{R_m(L_c)\sim(10^{16}-10^{18})}, using the upper value
\m{v_L^{\text{rms}}\sim0.1} in Eq.~(\ref{turb_vel_PT}), for
turbulence generated by first-order phase transitions. The
large magnetic Reynolds numbers, \m{R_m\gg
R_m^{\textrm{\tiny cr}}\approx60}
(for Kolmogorov turbulence) \cite{Brandenburg:2004jv},
indicate that
we are well within the regime where the SSD mechanism is
expected to operate.

\begin{figure}[h!]
\includegraphics[width=80mm,trim=20 20 20
20]{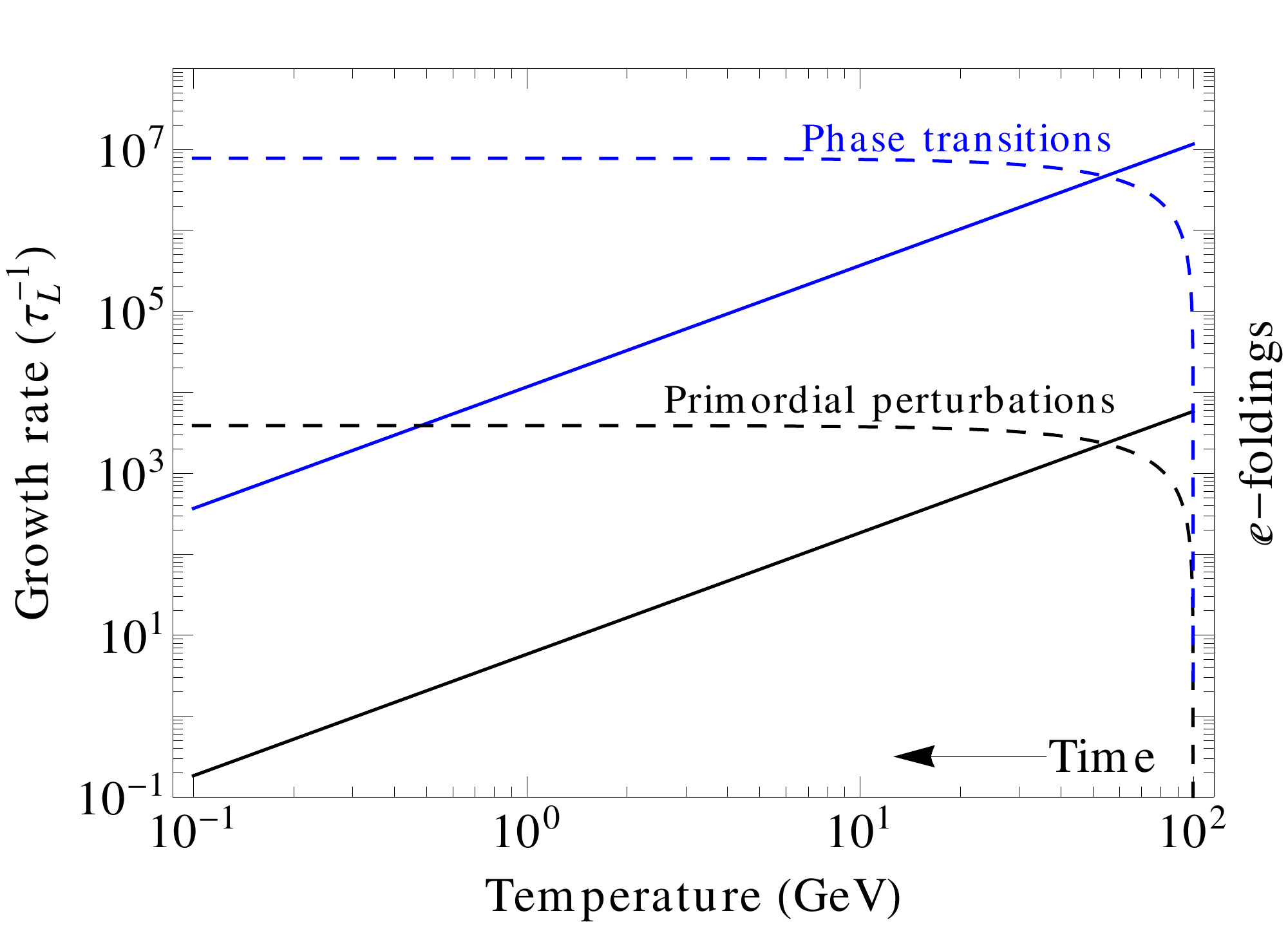}
\caption{\label{fig5} This figure shows the
SSD growth rate $\Gamma$ (solid) and number of $e$-foldings
\m{N=\int\Gamma(t)dt} (dashed) from the time of
the EW scale \m{T_{\text{EW}}\sim100~}GeV to the
QCD scale \m{T_{\text{QCD}}\simeq200}~MeV.
Recall that the magnetic seed field is amplified as
\m{B_{\text{rms}}\propto
\exp(\Gamma t)}. The growth
rate, shown here in units of $\tau_L^{-1}$, is determined
from Eq.~(\ref{gamma}) where
Kolmogorov turbulence \m{\vartheta=1/3} is assumed.
The growth rate and the number of $e$-foldings for
turbulence generated by the primordial density
perturbation and first-order phase transitions are shown in
(black) and (blue) respectively.}
\end{figure}

Figure~\ref{fig5} shows the magnetic field growth rate
$\Gamma$, where 
\m{B_{\text{rms}}\propto\exp(\Gamma t)}, which is determined
from the Kazantsev model of the SSD mechanism and given in
Eq.~(\ref{gamma}). The growth rate depends on the type of
turbulence, where \m{\vartheta=1/3,1/2}, applicable on the
inertial range
\m{l_{\text{diss}}<l<L} [cf. Eq.~(\ref{vel_turb_spec})], for
Kolmogorov and Burgers type turbulence respectively.
Here, we assume that the turbulence is of Kolmogorov type,
which is relevant for the subsonic velocity fluctuations
determined in this paper. Now, since $\Gamma$
varies in time, it will be useful to consider the number of
$e$-foldings given by \m{N\equiv\int\Gamma(t)\textrm{d}t}.
In Fig.~\ref{fig5}, the growth rate
is shown in units of the turnover rate of the largest eddy
$\tau_L^{-1}$ and $N(T)$ is shown
where $\Gamma(T)$ is integrated from 
\m{T=100}~GeV to temperature $T$. Since the magnetic field
grows as
\m{B_{\text{rms}}\propto\exp(\Gamma t)}, the number of
$e$-foldings gives
the total amplification factor. Due to very large Reynolds
numbers, the growth rate is initially very
large and the number of $e$-foldings quickly becomes large.
Hence, we find a very rapid increase in the field
strength and a huge amplification factor leading to rapid
saturation.

The phase of rapid exponential amplification comes to an end
when the magnetic energy becomes comparable to the kinetic
energy. In the radiation dominated epoch, this saturation
occurs when
\begin{equation}\label{saturation_RD}
 \langle \mbox{\boldmath
$B$}^2(\mbox{\boldmath $x$})\rangle 
\approx\frac{8}{45}\pi^3\varepsilon
g_*T^4\langle
\mbox{\boldmath
$v$}^2(\mbox{\boldmath
$x$})\rangle\,,
\end{equation}
where $g_*$ is the total number of effective relativistic
degrees of freedom, and the parameter
$\varepsilon$ quantifies the
saturation efficiency (see Sec.~\ref{SSD}). 
Without further dynamical evolution, the field strength will
only be diluted by the expansion \m{B\propto a^{-2}\propto
T^2}. Let us assume that the magnetic field becomes
saturated on the largest forcing scale $L_c$.
Thus, redshifted to present day values we find
\begin{equation}\label{saturation2} 
B^{\text{rms}}_0=a^2B_{\text{rms}}\approx\sqrt{\frac{8}{45}
\pi^3\varepsilon
g_*(T_*)}\,T_0^2
v_L^{\text{rms}}\,,
\end{equation}
where $T_0$ is the present day photon
temperature and $T_*$ is the radiation temperature at the
time of magnetic field amplification. With this simple
assumption, we find that in order to saturate magnetic
seed fields of strength \m{B_0^{\textrm{\tiny
seed}}\simeq(10^{-30}-10^{-20})~\textrm{nG}} up to
\m{\mathcal O(1)}~nG level
we require $e$-folding numbers \m{N\simeq46-70}. 

We can now estimate the saturated field strength from
Eq.~(\ref{saturation2}). For turbulence generated by the
primordial density perturbation, the kinetic
energy is given by the typical velocity fluctuations in
Eq.~(\ref{v_rms}). Hence, we find a saturated magnetic field
strength whose value today is
\m{B_0^{\text{rms}}\approx1\times10^{-9}\varepsilon^{1/2}}
~G. For turbulence generated by first-order phase
transitions, the kinetic energy is given by the velocity
fluctuations
in Eq.~(\ref{turb_vel_PT}) i.e. 
\m{v_L^{\text{rms}}\sim(10^{-4}-10^{-1})}. Hence,
we find the saturated field strengths in the range
\m{B_0^{\text{rms}}\approx(10^{-3}-1)\varepsilon^{1/2}
\mu\text {G}}. 

Here we note that the amplification up to saturation of 
magnetic seed fields from kinetic energy injected at first-order phase transitions
has been considered previously in the literature, see for example 
Ref.~\cite{B-phaseT} and references within. However, we believe it is important to point out that there is a well-defined 
and clearly described dynamo process that does the amplification, namely the small-scale dynamo 
theory provides the relevant framework for predictions regarding growth rates and saturation levels.

In the case where turbulence is generated by the primordial
density perturbation, the fluid forcing is continuous and
turbulence is fully developed throughout this epoch.
Hence, as we can see from Fig.~\ref{fig5}, the number of
$e$-foldings quickly become very large \m{N\gtrsim\mathcal
O(100)} and there is enough time in this epoch for tiny 
magnetic seed fields to saturate. However, for
turbulence generated by first-order phase transitions, the
fluid forcing is not continuous. In
Sec.~\ref{sec:turb_pdp}, we argue that for
\m{\alpha\lesssim1}, the time scale for the duration of
turbulence is approximately $\tau_L$
\cite{Kosowsky:2001xp}. Therefore,
the number of $e$-foldings $N\equiv\int\Gamma dt$ is roughly
given by \m{\Gamma/\tau_L^{-1}\approx R_e^{1/2}\gg1} (see
Eq.~(\ref{gamma}) assuming a Kolmogorov spectrum). Hence,
the
magnetic fields can easily become saturated in the time
scale of the phase transition. 

The saturated field strengths are very strong. However, the
scales over which we expect saturation are very small.
We expect saturation on scales from the damping scale
$l_D$ up to the largest forcing
scale $L_c$. For a Kolmogorov spectrum, where most power
resides on the largest scale, we can identify $L_c$ as the
comoving coherence
length $\lambda_c$ of the magnetic field. For turbulence
generated by the primordial density perturbation, the
largest forcing scale in the epoch considered is
\m{L_c} at \m{T\simeq0.2\text{GeV}}
i.e. \m{\lambda_c\sim 10^{-5}\text{pc}} (see
Fig.~\ref{fig2}). For turbulence generated by first-order
phase transitions, the length scales on which
saturation is expected depends on the exact time of
the phase transition. The basic constraint on
the coherence length is the horizon size at the time of the
phase transition; \m{\lambda_c\simeq0.1}~pc
and \m{\lambda_c\simeq10^{-4}}~pc for
the QCD and EW phase transitions respectively.


To obtain larger coherence lengths, the
magnetic field would need to be amplified and saturated at a
later time when the Hubble horizon is larger and turbulence
develops on larger scales. Unfortunately, the velocity
perturbations generated by the primordial density
perturbation are efficiently damped
below \m{T\simeq0.2}~GeV and therefore do not lead to a
state of fully developed turbulence. Without turbulence
there is no SSD action and therefore no amplification of
primordial magnetic fields. For the SSD mechanism to be
effective at a later time in the RD era, a different
mechanism which injects kinetic energy into the plasma is
required.

\subsection{Subsequent evolution}\label{sec:sub-evo}

In the previous section we considered the amplification of magnetic seed fields 
at the time when turbulent kinetic energy is injected into the primordial plasma.
The integral scale of the magnetic field, which is amplified up to equipartition
with the kinetic energy, is determined by the scale of 
the injected turbulence. In this section we consider the cosmological evolution of the
magnetic field strengths and coherence lengths from the time 
after the injection of turbulence to the present time. We follow the works of
Refs.~\cite{Jedamzik:1996wp,Banerjee:2004df,Jedamzik:2010cy,
Campanelli_free-turb,Durrer:2013pga} for the growth of the coherence length and the
damping of turbulence in this \emph{subsequent} regime in order to determine the final
magnetic field strengths and coherence lengths. The most important result from such works is that magnetic
fields generated in early epochs survive through viscous
and free-streaming regimes, unlike velocity
perturbations which are efficiently damped (see
Sec.~\ref{Damping}).

In the \emph{turbulent} regimes, strong magnetic
fields on small scales drive turbulence in the plasma up
to equipartition. The turbulence removes power on small
scales thereby increasing the correlation length and
reducing the field strength. This turbulent magnetohydrodynamic (MHD) effect,
free turbulent decay, depends on the type of turbulence
generated and on whether or not the magnetic field is
helical~\cite{Banerjee:2004df,Durrer:2013pga}. For
nonhelical fields, the growth of the coherence length is
purely due to the dissipation of power on
small scales. In the helical case, there is an \emph{inverse
cascade} effect where the power on larger scales
grows~\cite{Brandenburg:1996fc,Banerjee:2004df}.

Let us only consider nonhelical magnetic fields in
turbulent regimes. The growth of $\lambda_c$ is a power law
in time with an index that depends on the magnetic field
spectrum $n_B$. For magnetic fields generated by causal
processes, e.g. phase transitions, the index is \m{n_B=2}
\cite{CausalBfield}. In this case 
\m{\lambda_c\sim t^{2/7},t^{2/5}} and the
magnetic field strength at the scale $\lambda_c$ evolves as
\m{a^2B_{\text{rms}}\sim t^{-5/7},t^{-3/5}}
for incompressible (Kolmogorov)
and compressible (Burgers) type turbulence
respectively~\cite{Banerjee:2004df,Durrer:2013pga,
Jedamzik:2010cy,Campanelli_free-turb}. Hence, the evolution
of the field strength and coherence length up to the time of
recombination is determined by the
relation
\begin{equation}\label{B_evol1}
 a^2B_{\text{rms}}\sim \lambda_c^{-n}\,,
\end{equation}
where \m{n=\frac52,\frac32} for Kolmogorov and Burgers
turbulence
respectively~\cite{Banerjee:2004df,Durrer:2013pga,
Jedamzik:2010cy,Campanelli_free-turb}.

Besides the evolution of magnetic fields in the 
turbulent regimes, in the RD era, there are also epochs
of viscous damping and free-streaming. In a magnetized
plasma, there are different modes in which magnetic energy
can be stored; fast, slow and Alfv\'{e}n modes, as opposed
to only the acoustic mode in the case of velocity
fluctuations. The fast magnetosonic mode decays in the same
manner as the acoustic mode due to particle diffusion, see
Eq.~(\ref{l_Damping}). However, the slow and Alfv\'{e}n
modes evolve differently and can become
overdamped~\cite{Jedamzik:1996wp}. Hence, the magnetic
energy stored in these modes becomes frozen-in. The
overdamping depends on the scales and magnetic field
strength. Therefore, magnetic fields survive through viscous
and free-streaming regimes, which is in contrast to
turbulent velocity fluctuations that become efficiently
damped. In the viscous damping and free-streaming regimes,
the evolution of \m{a^2B_{\text{rms}}} and
$\lambda_c$ is halted until free turbulent decay begins
again~\cite{Banerjee:2004df,Durrer:2013pga}. 
The evolution due to free turbulent decay terminates when
the correlation length and field strength end on the line,
in the \m{\{a^2B_{\text{rms}},\lambda_c\}} plane, given
by~\cite{Banerjee:2004df,Durrer:2013pga}
\begin{equation}\label{Blambda}
 a^2B_{\text{rms}}\simeq
10^{-8}\frac{\lambda_c}{\text{Mpc}}~\text{G}\,.
\end{equation}
This line corresponds to the largest eddies being processed
at recombination
\m{1/(a_{\text{rec}}H_{\text{rec}})\simeq\lambda_c/v_A}
with $v_A$ the Alfv\'{e}n speed~\cite{Durrer:2013pga}.
In the matter dominated Universe, there is no further
evolution of the magnetic field correlation length
$\lambda_c$, although strong fields on small scales can
drive turbulence at much later times in the intergalactic
medium and restore the turbulent
decay~\cite{Durrer:2013pga}. In this sense,
Eq.~(\ref{Blambda}) becomes an upper bound on the present
day magnetic field strength. The difference in evolution
of the magnetic modes, in contrast to acoustic modes, means
that fields generated and amplified in the radiation
dominated era can survive to the present day.

As seen in the previous subsection, turbulence generated by
the primordial density perturbation
can amplify tiny magnetic seed fields through the SSD
mechanism to values of order
\m{a^2B_{\text{rms}}\sim1\varepsilon^{1/2}}~nG 
on scales at most \m{\lambda_c\sim10^{-5}}~pc. 
From eqs.~(\ref{B_evol1}) and (\ref{Blambda}), these fields
would evolve to
\m{a^2B_{\text{rms}}\sim10^{-6}\varepsilon^{1/2}}~nG on
scales \m{\lambda_c\sim10^{-1}}~pc. 
Since the primordial density perturbation is necessarily present
for structure formation, such fields are guaranteed by the 
SSD mechanism and can play an important role in structure formation.
Unfortunately, these
fields are too weak on too short scales to explain the
Fermi observations of TeV Blazars~\cite{Neronov:1900zz}.
However, turbulence generated by
first-order phase transitions can amplify magnetic seed
fields to values of order
\m{a^2B_{\text{rms}}\sim(10^{-3}-1)\varepsilon^{1/2}}
~$\mu$G on scales
\m{\lambda_c\sim(10^{-4}-10^{-1})}~pc. These initial field
configurations will evolve to 
\m{a^2B_{\text{rms}}\sim(10^{-6}-10^{-3})\varepsilon^{1/2}}
~nG on scales \m{\lambda_c\sim(10^{-1}-10^{2})}~pc. Such
fields are strong enough to explain the
Fermi observations of TeV Blazars~\cite{Neronov:1900zz}.

\section{Summary}\label{sec:sum}

In this paper we have identified two mechanisms that
generate turbulence in the radiation dominated Universe.
The two mechanisms inject kinetic energy into the
primordial plasma at times when the kinetic Reynolds
numbers are very large \m{R_e\gg1}. With the injection of
kinetic energy, which is determined by
\m{v_L^{\text{rms}}}, and large Reynolds numbers, a state a
fully developed turbulence is expected on the
\textit{inertial range}: \m{l_{\text{diss},c}<l_c<L_c},
where
\m{L_c=v_L^{\text{rms}}/aH} is the largest forcing scale and
\m{l_{\text{diss},c}\sim L_cR_e(L)^{-3/4}} is the
dissipative scale (for Kolmogorov turbulence). The two
mechanisms for generating velocity fluctuations are:
by the primordial density perturbation and bubble collisions
during first-order phase
transitions. In Sec.~\ref{SSDamp}, we have seen that
turbulence is inevitably generated by the primordial density
perturbation prior to the QCD scale \m{T\gtrsim200}~MeV. In
Sec.~\ref{sec:PT}, we investigated the possibility of
turbulence generated at first-order
phase transitions between the electroweak (EW) and the QCD
scales \m{200~\text{MeV}\lesssim T\lesssim100}~GeV. 
For the generation of turbulence at first-order
phase transitions, we have simply followed the same
assumptions made in the literature regarding the dynamics
of the phase transitions \cite{Kamionkowski:1993fg,
Kosowsky:2001xp, Caprini:2006jb}.

The turbulence generated can amplify tiny magnetic seed
fields through the small-scale dynamo (SSD) mechanism. 
If we assume that the plasma is already weakly magnetized
by the time of the EW scale, then SSD amplification will
inevitable occur. The rapid amplification ends at saturation
when the magnetic and kinetic energies are in approximate
equipartition
\m{E_{\text{M}}/E_{\text{kin}}\approx\varepsilon}, where the
efficiency of the mechanism is characterized by
$\varepsilon$. The saturation efficiency parameter
$\varepsilon$, which is
determined numerically, varies depending on the type of
forcing. For rotational modes \m{\varepsilon\approx1},
whereas the saturation level
is lower for compressive modes
\m{\varepsilon\sim10^{-3}-10^{-4}}~\cite{Federrath:2011qz}.
We note that these numerical studies were carried out for
\m{P_m\approx2}, further numerical work is required to
establish the saturation level for large Prandtl numbers. 

In this paper, we show that, even
for tiny seed fields of strengths \m{B_0^{\textrm{\tiny
seed}}\simeq(10^{-30}-10^{-20})~\textrm{nG}}, 
the SSD mechanism can operate for a long enough period
of time and be efficient enough to amplify such fields to
saturation. The magnetic field strength saturates at
\m{a^2B_{\text{rms}}\sim1\varepsilon^{1/2}}~nG 
on scales at most \m{\lambda_c\sim10^{-5}}~pc for
turbulence generated by the primordial density perturbation.
Such fields, assumed to be nonhelical, evolve to
\m{B_0^{\text{rms}}\sim10^{-6}\varepsilon^{1/2}}~nG on
scales \m{\lambda_c\sim10^{-1}}~pc due to free turbulent
decay. For turbulence generated by first-order phase
transitions, the SSD mechanism can be even more effective,
since the
turbulent velocities can be quite large compared
to those generated by the primordial density perturbation.
We show that the mechanism can amplify magnetic fields to
strengths
\m{a^2B_{\text{rms}}\sim(10^{-3}-1)\varepsilon^{1/2}}~$\mu$G
on scales \m{\lambda_c\sim(10^{-4}-10^{-1})}~pc. These
initial field configurations evolve to 
\m{a^2B_{\text{rms}}\sim(10^{-6}-10^{-3})\varepsilon^{1/2}}
~nG on scales
\m{\lambda_c\sim(10^{-1}-10^{2})}~pc due to free turbulent
decay.

Unfortunately, the damping of velocity perturbations
due to neutrino diffusion inhibits turbulence
from developing below the QCD scale
\m{T_{\text{QCD}}\simeq200}~MeV (unless there is an
injection of kinetic energy from a first-order phase
transition prior to neutrino decoupling at
\m{T\gtrsim2.6}~MeV, see Fig.~\ref{fig2}). Hence, it is
difficult to generate turbulence from these mechanisms on
larger length scales than \m{l_c\sim10^{-1}}~pc. Without
turbulence there is no SSD action and therefore no
amplification of primordial magnetic fields on larger
scales.

Although turbulent velocities are completely erased
in viscous and free-streaming regimes,
magnetic fields are overdamped and can survive to the
present day. Such fields would fill the voids in the large
scale structure and provide the seeds for magnetic fields
generated by structure formation and galactic dynamo.
Unfortunately, the saturated field strengths due to
turbulence generated by the primordial density perturbation
are too weak on too short scales in the voids of the large
scale structure to explain the Fermi observations of TeV
Blazars~\cite{Neronov:1900zz}. However, the field strengths 
obtained due to turbulence generated by first-order phase
transitions are strong enough to explain such
observations.

\section{Discussion and conclusion}\label{sec:conc}

The discrepancy between theoretically generated and observed
magnetic fields in the Universe needs explaining. The
galactic dynamo can be a very effective
mechanism at producing the $\mu$G fields observed in spiral
galaxies~\cite{BFields_galax}. However, strong fields in
young galaxies, clusters and superclusters of galaxies and
in the intergalactic medium require further
explanation~\cite{BFields_Bernet,BFields-Clusters,
BFields_SClusters,greenpeas,Neronov:1900zz}. As
noted in a number of numerical and analytical works, the
rapid amplification of magnetic seed fields can occur due
to the turbulent motions of the conducting plasma. This
small-scale dynamo (SSD) mechanism is believed to play a crucial
role in the formation of large magnetic fields in a number
of astronomical settings, from stars to galaxies
and the intergalactic
medium~\cite{BFields_Beck,Schleicher:2010ph,Latif:2012aq,
Banerjee:2012ch, Schober:2013aoa}. For these
settings, the turbulent motions arise from
gravitational collapse, accretion and supernovae
explosions. Hence, the SSD mechanism can be highly effective
at magnetizing structures in the early Universe. However,
the large field strengths apparently observed
in the voids of the large-scale structure
\cite{Neronov:1900zz} still require an explanation.

Magnetic seed fields will
almost certainly be
generated at some level in the early Universe through a
variety of mechanisms. 
Such mechanisms include inflation~\cite{Turner:PMF}, phase
transitions~\cite{B-phaseT} and the
Harrison mechanism through the generation of
vorticity~\cite{Harrison}. The SSD mechanism for the
amplification of such seed fields
could play an important role for the explanation of the
observed large magnetic fields throughout the Universe.
In this paper we have demonstrated that the conditions
necessary for such turbulent amplification arise in the
radiation dominated Universe before the onset of structure
formation. We have shown that significant turbulence is
generated in this early epoch by at least two mechanisms;
velocity perturbations generated by the primordial density
perturbation and bubble collisions in first-order phase
transitions. 

%
%

Turbulent plasma motions arise inevitably from perturbations
of the gravitational potential. The continuous production of
velocity perturbations upon horizon entry of primordial
density modes, act as a continuous forcing of the fluid on
the largest scales. Therefore, in regimes of large Reynolds
numbers, a state of stationary fully developed turbulence is
expected. Turbulent flow can be triggered, for example, by
thermal fluctuations on very small scales~\cite{Tsuge:1974}.
Turbulence can also be injected into the
plasma by bubble collisions during first-order phase
transitions. Although the kinetic energy injection  
occurs only for the duration of the phase transition, we
argue, following
Refs.~\cite{Kosowsky:2001xp,Caprini:2006jb}, that a state of
fully developed turbulence is also
expected from this mechanism.

Once fully developed turbulence is established, the
Kazantsev model of the SSD mechanism can be
used to estimate the magnetic field growth rate. We have
demonstrated that the Prandtl numbers are very large
in the regime considered. Thus, the results from the
Kazantsev theory for \m{P_m\gg1} are applicable.
The analytical work shows that the magnetic field growth
rate depends
on the kinetic Reynolds numbers
\cite{SSD-Re_Pm,Schekochihin:2001pj},
which are very large in our case. We have shown that, for
both models of turbulence, the amplification is strong
enough for small magnetic seed fields to reach a
saturated state. The saturated state is given by the
approximate equipartition between magnetic and kinetic
energy \m{E_{\text{M}}/E_{\text{kin}}\approx\varepsilon},
where the parameter $\varepsilon$ characterizes the
efficiency of the mechanism. 

We note that numerical studies at Prandtl numbers
\m{P_m\approx2} indicate that the SSD
mechanism is more efficient for rotational modes, where the
saturation efficiency $\varepsilon$ is close to
unity~\cite{Federrath:2011qz}. Whereas the saturation level
is lower for compressive modes
\m{\varepsilon\sim10^{-3}-10^{-4}}~\cite{Federrath:2011qz}.
However, further numerical work is required to establish the
saturation level for larger Prandtl numbers and smaller Mach
numbers relevant to our settings. We also note that,
although only longitudinal velocity modes are generated by
first-order primordial density perturbations, rotational
modes are generated at second order in cosmological
perturbations~\cite{Lewis:2004kg,
Lu:2007cj, Lu:2008ju, Christopherson:2010dw,
Ichiki:2011ah}. Also, there is no reason not to expect
rotational modes generated by first-order phase
transitions. In any case, since the Reynolds numbers are so
large, nonlinear interactions can play a role leading to a
state of fully developed turbulence with both rotational
and longitudinal modes. In particular, we expect that, below
the integral scale, Kolmogorov type turbulence is
established. But we stress that the SSD
mechanism works independently of the type of
turbulence~\cite{Gruzinov:1996gm,Schekochihin:2001pj,
SSD-Re_Pm}. Indeed, even purely irrotational turbulence can
still drive a small-scale
dynamo~\cite{Schekochihin:2001pj,SSD-Re_Pm,Federrath:2011qz}
.
Hence, the efficient amplification of magnetic fields seems
unavoidable, leading to a strongly
magnetized early Universe prior to structure formation.

For the two mechanisms of turbulence investigated
in this paper, we calculated the saturated field strengths
and their subsequent evolution up to the
present day. We note that although turbulence is
completely erased in viscous and free-streaming regimes,
magnetic fields are overdamped and can survive to the
present day. Therefore, the most
important epochs of evolution are due to free turbulent
decay. This turbulent MHD effect decreases the field
strength and increases the coherence
length in nonhelical
fields~\cite{Banerjee:2004df,Durrer:2013pga}. From the
turbulence
generated by the primordial density perturbation we found
\m{B_0^{\text{rms}}\sim10^{-6}\varepsilon^{1/2}}~nG on
scales \m{\lambda_c\sim10^{-1}}~pc. Unfortunately, even for
a high efficiency factor \m{\varepsilon\sim1},
these fields are too weak on too short scales to explain
the Fermi observations of TeV Blazars~\cite{Neronov:1900zz}.
From the turbulence
generated by first-order phase transitions, we found 
\m{B_0^{\text{rms}}\sim(10^{-6}-10^{-3})\varepsilon^{1/2}}
~nG on scales
\m{\lambda_c\sim(10^{-1}-10^{2})}~pc. Such fields are
strong enough to explain the apparent observations of 
intergalactic magnetic fields suggested by the 
Fermi results~\cite{Neronov:1900zz}. Thus, in this paper we
have demonstrated that the
conditions are right for the efficient amplification
of magnetic fields via the small-scale dynamo. The mechanism
generates large field strengths, albeit on very small
scales, which could explain observations of magnetic fields
in the voids of the large-scale structure and
have an impact on early structure formation.

\begin{acknowledgments}
We wish to thank the anonymous referees for their constructive comments.
This work was supported by the Deutsche Forschungsgemeinschaft through the collaborative 
research centre SFB 676, by the \emph{Helmholtz Alliance for Astroparticle Phyics} (HAP) 
funded by the Initiative and Networking
Fund of the Helmholtz Association. DRGS thanks for funding by the German Science 
Foundation (DFG) via the SFB 963/1 \emph{Astrophysical Flow Instabilities and Turbulence} (project A12).
\end{acknowledgments}


\begin{thebibliography}{50}%

\makeatletter
\providecommand \@ifxundefined [1]{%
 \ifx #1\undefined \expandafter \@firstoftwo
 \else \expandafter \@secondoftwo
\fi
}%
\providecommand \@ifnum [1]{%
 \ifnum #1\expandafter \@firstoftwo
 \else \expandafter \@secondoftwo
\fi
}%
\providecommand \enquote [1]{``#1''}%
\providecommand \bibnamefont  [1]{#1}%
\providecommand \bibfnamefont [1]{#1}%
\providecommand \citenamefont [1]{#1}%
\providecommand\href[0]{\@sanitize\@href}%
\providecommand\@href[1]{\endgroup\@@startlink{#1}\endgroup\@@href}%
\providecommand\@@href[1]{#1\@@endlink}%
\providecommand \@sanitize [0]{\begingroup\catcode`\&12\catcode`\#12\relax}%
\@ifxundefined \pdfoutput {\@firstoftwo}{%
 \@ifnum{\z@=\pdfoutput}{\@firstoftwo}{\@secondoftwo}%
}{%
 \providecommand\@@startlink[1]{\leavevmode}%
 \providecommand\@@endlink[0]{}%
}{%
 \providecommand\@@startlink[1]{%
  \leavevmode
  \pdfstartlink
   attr{/Border[0 0 1 ]/H/I/C[0 1 1]}%
   user{/Subtype/Link/A<</Type/Action/S/URI/URI(#1)>>}%
  \relax
 }%
 \providecommand\@@endlink[0]{\pdfendlink}%
}%
\providecommand \url  [0]{\begingroup\@sanitize \@url }%
\providecommand \@url [1]{\endgroup\@href {#1}{\urlprefix}}%
\providecommand \urlprefix [0]{URL }%
\providecommand \Eprint[0]{\href }%
\@ifxundefined \urlstyle {%
  \providecommand \doi [1]{doi:\discretionary{}{}{}#1}%
}{%
  \providecommand \doi [0]{doi:\discretionary{}{}{}\begingroup
  \urlstyle{rm}\Url }%
}%
\providecommand \doibase [0]{http://dx.doi.org/}%
\providecommand \Doi[1]{\href{\doibase#1}}%
\providecommand \bibAnnote [3]{%
  \BibitemShut{#1}%
  \begin{quotation}\noindent
    \textsc{Key:}\ #2\\\textsc{Annotation:}\ #3%
  \end{quotation}%
}%
\providecommand \bibAnnoteFile [2]{%
  \IfFileExists{#2}{\bibAnnote {#1} {#2} {\input{#2}}}{}%
}%
\providecommand \typeout [0]{\immediate \write \m@ne }%
\providecommand \selectlanguage [0]{\@gobble}%
\providecommand \bibinfo [0]{\@secondoftwo}%
\providecommand \bibfield [0]{\@secondoftwo}%
\providecommand \translation [1]{[#1]}%
\providecommand \BibitemOpen[0]{}%
\providecommand \bibitemStop [0]{}%
\providecommand \bibitemNoStop [0]{.\EOS\space}%
\providecommand \EOS [0]{\spacefactor3000\relax}%
\providecommand \BibitemShut [1]{\csname bibitem#1\endcsname}%
\bibitem{BFields_galax}%
  \BibitemOpen
  \bibfield{author}{%
  \bibinfo {author} {\bibfnamefont{R.}~\bibnamefont{Beck}},\ }%
  \bibfield{journal}{%
  \Doi{10.1007/s11214-011-9782-z}{\bibinfo {journal} {Space Science Reviews}}\
  }%
  \textbf{\bibinfo {volume} {166}},\ \bibinfo {pages} {215} (\bibinfo {year}
  {2012})%
  \bibAnnoteFile{NoStop}{BFields_galax}%
\bibitem{BFields_Bernet}%
  \BibitemOpen
  \bibfield{author}{%
  \bibinfo {author} {\bibfnamefont{M.~L.}\ \bibnamefont{Bernet}}, \bibinfo
  {author} {\bibfnamefont{F.}~\bibnamefont{Miniati}}, \bibinfo {author}
  {\bibfnamefont{S.~J.}\ \bibnamefont{Lilly}}, \bibinfo {author}
  {\bibfnamefont{P.~P.}\ \bibnamefont{Kronberg}},\ and\ \bibinfo {author}
  {\bibfnamefont{M.}~\bibnamefont{Dessauges-Zavadsky}},\ }%
  \bibfield{journal}{%
  \Doi{10.1038/nature07105}{\bibinfo {journal} {Nature}}\ }%
  \textbf{\bibinfo {volume} {454}},\ \bibinfo {pages} {302} (\bibinfo {year}
  {2008}),\ \Eprint{http://arxiv.org/abs/0807.3347}{arXiv:0807.3347}%
  \bibAnnoteFile{NoStop}{BFields_Bernet}%
\bibitem{BFields-Clusters}%
  \BibitemOpen
  \bibfield{author}{%
  \bibinfo {author} {\bibfnamefont{L.}~\bibnamefont{Feretti}}, \bibinfo
  {author} {\bibfnamefont{G.}~\bibnamefont{Giovannini}}, \bibinfo {author}
  {\bibfnamefont{F.}~\bibnamefont{Govoni}},\ and\ \bibinfo {author}
  {\bibfnamefont{M.}~\bibnamefont{Murgia}},\ }%
  \bibfield{journal}{%
  \Doi{10.1007/s00159-012-0054-z}{\bibinfo {journal} {A\&AR}}\ }%
  \textbf{\bibinfo {volume} {20}},\ \bibinfo {pages} {54} (\bibinfo {year}
  {2012}),\ \Eprint{http://arxiv.org/abs/1205.1919}{arXiv:1205.1919
  [astro-ph.CO]}%
  \bibAnnoteFile{NoStop}{BFields-Clusters}%
\bibitem{BFields_SClusters}%
  \BibitemOpen
  \bibfield{author}{%
  \bibinfo {author} {\bibfnamefont{Y.}~\bibnamefont{Xu}}, \bibinfo {author}
  {\bibfnamefont{P.~P.}\ \bibnamefont{Kronberg}}, \bibinfo {author}
  {\bibfnamefont{S.}~\bibnamefont{Habib}},\ and\ \bibinfo {author}
  {\bibfnamefont{Q.~W.}\ \bibnamefont{Dufton}},\ }%
  \bibfield{journal}{%
  \Doi{10.1086/498336}{\bibinfo {journal} {AstrophysJ}}\ }%
  \textbf{\bibinfo {volume} {637}},\ \bibinfo {pages} {19} (\bibinfo {year}
  {2006}),\
  \Eprint{http://arxiv.org/abs/arXiv:astro-ph/0509826}{arXiv:astro-ph/0509826}%
  \bibAnnoteFile{NoStop}{BFields_SClusters}%
\bibitem{greenpeas}%
  \BibitemOpen
  \bibfield{author}{%
  \bibinfo {author} {\bibfnamefont{S.}~\bibnamefont{{Chakraborti}}}, \bibinfo
  {author} {\bibfnamefont{N.}~\bibnamefont{{Yadav}}}, \bibinfo {author}
  {\bibfnamefont{C.}~\bibnamefont{{Cardamone}}},\ and\ \bibinfo {author}
  {\bibfnamefont{A.}~\bibnamefont{{Ray}}},\ }%
  \bibfield{journal}{%
  \Doi{10.1088/2041-8205/746/1/L6}{\bibinfo {journal} {Astrophys.\, J.\,}}\ }%
  \textbf{\bibinfo {volume} {746}},\ \bibinfo {eid} {L6} (\bibinfo {month}
  {Feb.}\ \bibinfo {year} {2012}),\
  \Eprint{http://arxiv.org/abs/1110.3312}{arXiv:1110.3312 [astro-ph.CO]}%
  \bibAnnoteFile{NoStop}{greenpeas}%
\bibitem{Neronov:1900zz}%
  \BibitemOpen
  \bibfield{author}{%
  \bibinfo {author} {\bibfnamefont{A.}~\bibnamefont{Neronov}}\ and\ \bibinfo
  {author} {\bibfnamefont{I.}~\bibnamefont{Vovk}},\ }%
  \bibfield{journal}{%
  \Doi{10.1126/science.1184192}{\bibinfo {journal} {Science}}\ }%
  \textbf{\bibinfo {volume} {328}},\ \bibinfo {pages} {73} (\bibinfo {year}
  {2010}),\ \Eprint{http://arxiv.org/abs/1006.3504}{arXiv:1006.3504
  [astro-ph.HE]}%
  \bibAnnoteFile{NoStop}{Neronov:1900zz}%
\bibitem{Broderick_etal_2012}%
  \BibitemOpen
  \bibfield{author}{%
  \bibinfo {author} {\bibfnamefont{A.~E.}\ \bibnamefont{{Broderick}}}, \bibinfo
  {author} {\bibfnamefont{P.}~\bibnamefont{{Chang}}},\ and\ \bibinfo {author}
  {\bibfnamefont{C.}~\bibnamefont{{Pfrommer}}},\ }%
  \bibfield{journal}{%
  \Doi{10.1088/0004-637X/752/1/22}{\bibinfo {journal} {Astrophys.\, J.\,}}\ }%
  \textbf{\bibinfo {volume} {752}},\ \bibinfo {eid} {22} (\bibinfo {month}
  {Jun.}\ \bibinfo {year} {2012}),\
  \Eprint{http://arxiv.org/abs/1106.5494}{arXiv:1106.5494 [astro-ph.CO]}%
  \bibAnnoteFile{NoStop}{Broderick_etal_2012}%
\bibitem{Schlickeiser_etal_2012}%
  \BibitemOpen
  \bibfield{author}{%
  \bibinfo {author} {\bibfnamefont{R.}~\bibnamefont{{Schlickeiser}}}, \bibinfo
  {author} {\bibfnamefont{D.}~\bibnamefont{{Ibscher}}},\ and\ \bibinfo {author}
  {\bibfnamefont{M.}~\bibnamefont{{Supsar}}},\ }%
  \bibfield{journal}{%
  \Doi{10.1088/0004-637X/758/2/102}{\bibinfo {journal} {Astrophys.\, J.\,}}\ }%
  \textbf{\bibinfo {volume} {758}},\ \bibinfo {eid} {102} (\bibinfo {month}
  {Oct.}\ \bibinfo {year} {2012})%
  \bibAnnoteFile{NoStop}{Schlickeiser_etal_2012}%
\bibitem{Turner:PMF}%
  \BibitemOpen
  \bibfield{author}{%
  \bibinfo {author} {\bibfnamefont{M.~S.}\ \bibnamefont{Turner}}\ and\ \bibinfo
  {author} {\bibfnamefont{L.~M.}\ \bibnamefont{Widrow}},\ }%
  \bibfield{journal}{%
  \Doi{10.1103/PhysRevD.37.2743}{\bibinfo {journal} {Phys.Rev.}}\ }%
  \textbf{\bibinfo {volume} {D37}},\ \bibinfo {pages} {2743} (\bibinfo {year}
  {1988})%
  \bibAnnoteFile{NoStop}{Turner:PMF}%
\bibitem{B-phaseT}%
  \BibitemOpen
  \bibfield{author}{%
  \bibinfo {author} {\bibfnamefont{G.}~\bibnamefont{Sigl}}, \bibinfo {author}
  {\bibfnamefont{A.~V.}\ \bibnamefont{Olinto}},\ and\ \bibinfo {author}
  {\bibfnamefont{K.}~\bibnamefont{Jedamzik}},\ }%
  \bibfield{journal}{%
  \Doi{10.1103/PhysRevD.55.4582}{\bibinfo {journal} {Phys.Rev.}}\ }%
  \textbf{\bibinfo {volume} {D55}},\ \bibinfo {pages} {4582} (\bibinfo {year}
  {1997}),\
  \Eprint{http://arxiv.org/abs/astro-ph/9610201}{arXiv:astro-ph/9610201
  [astro-ph]}%
  \bibAnnoteFile{NoStop}{B-phaseT}%
\bibitem{Dolgov:2001nv}%
  \BibitemOpen
  \bibfield{author}{%
  \bibinfo {author} {\bibfnamefont{A.~D.}\ \bibnamefont{Dolgov}}\ and\ \bibinfo
  {author} {\bibfnamefont{D.}~\bibnamefont{Grasso}},\ }%
  \bibfield{journal}{%
  \Doi{10.1103/PhysRevLett.88.011301}{\bibinfo {journal} {Phys.Rev.Lett.}}\ }%
  \textbf{\bibinfo {volume} {88}},\ \bibinfo {pages} {011301} (\bibinfo {year}
  {2001}),\
  \Eprint{http://arxiv.org/abs/astro-ph/0106154}{arXiv:astro-ph/0106154
  [astro-ph]}%
  \bibAnnoteFile{NoStop}{Dolgov:2001nv}%
\bibitem{Matarrese-B}%
  \BibitemOpen
  \bibfield{author}{%
  \bibinfo {author} {\bibfnamefont{S.}~\bibnamefont{Matarrese}}, \bibinfo
  {author} {\bibfnamefont{S.}~\bibnamefont{Mollerach}}, \bibinfo {author}
  {\bibfnamefont{A.}~\bibnamefont{Notari}},\ and\ \bibinfo {author}
  {\bibfnamefont{A.}~\bibnamefont{Riotto}},\ }%
  \bibfield{journal}{%
  \Doi{10.1103/PhysRevD.71.043502}{\bibinfo {journal} {Phys.Rev.}}\ }%
  \textbf{\bibinfo {volume} {D71}},\ \bibinfo {pages} {043502} (\bibinfo {year}
  {2005}),\
  \Eprint{http://arxiv.org/abs/astro-ph/0410687}{arXiv:astro-ph/0410687
  [astro-ph]}%
  \bibAnnoteFile{NoStop}{Matarrese-B}%
\bibitem{Takahashi:2005nd}%
  \BibitemOpen
  \bibfield{author}{%
  \bibinfo {author} {\bibfnamefont{K.}~\bibnamefont{Takahashi}}, \bibinfo
  {author} {\bibfnamefont{K.}~\bibnamefont{Ichiki}}, \bibinfo {author}
  {\bibfnamefont{H.}~\bibnamefont{Ohno}},\ and\ \bibinfo {author}
  {\bibfnamefont{H.}~\bibnamefont{Hanayama}},\ }%
  \bibfield{journal}{%
  \Doi{10.1103/PhysRevLett.95.121301}{\bibinfo {journal} {Phys.Rev.Lett.}}\ }%
  \textbf{\bibinfo {volume} {95}},\ \bibinfo {pages} {121301} (\bibinfo {year}
  {2005}),\
  \Eprint{http://arxiv.org/abs/astro-ph/0502283}{arXiv:astro-ph/0502283
  [astro-ph]}%
  \bibAnnoteFile{NoStop}{Takahashi:2005nd}%
\bibitem{Kobayashi:2007wd}%
  \BibitemOpen
  \bibfield{author}{%
  \bibinfo {author} {\bibfnamefont{T.}~\bibnamefont{Kobayashi}}, \bibinfo
  {author} {\bibfnamefont{R.}~\bibnamefont{Maartens}}, \bibinfo {author}
  {\bibfnamefont{T.}~\bibnamefont{Shiromizu}},\ and\ \bibinfo {author}
  {\bibfnamefont{K.}~\bibnamefont{Takahashi}},\ }%
  \bibfield{journal}{%
  \Doi{10.1103/PhysRevD.75.103501}{\bibinfo {journal} {Phys.Rev.}}\ }%
  \textbf{\bibinfo {volume} {D75}},\ \bibinfo {pages} {103501} (\bibinfo {year}
  {2007}),\
  \Eprint{http://arxiv.org/abs/astro-ph/0701596}{arXiv:astro-ph/0701596
  [astro-ph]}%
  \bibAnnoteFile{NoStop}{Kobayashi:2007wd}%
\bibitem{Fenu:2010kh}%
  \BibitemOpen
  \bibfield{author}{%
  \bibinfo {author} {\bibfnamefont{E.}~\bibnamefont{Fenu}}, \bibinfo {author}
  {\bibfnamefont{C.}~\bibnamefont{Pitrou}},\ and\ \bibinfo {author}
  {\bibfnamefont{R.}~\bibnamefont{Maartens}},\ }%
  \bibfield{journal}{%
  \Doi{10.1111/j.1365-2966.2011.18554.x}{\bibinfo {journal}
  {Mon.Not.Roy.Astron.Soc.}}\ }%
  \textbf{\bibinfo {volume} {414}},\ \bibinfo {pages} {2354} (\bibinfo {year}
  {2011}),\ \Eprint{http://arxiv.org/abs/1012.2958}{arXiv:1012.2958
  [astro-ph.CO]}%
  \bibAnnoteFile{NoStop}{Fenu:2010kh}%
\bibitem{Harrison}%
  \BibitemOpen
  \bibfield{author}{%
  \bibinfo {author} {\bibfnamefont{E.}~\bibnamefont{Harrison}},\ }%
  \bibfield{journal}{%
  \bibinfo {journal} {Mon.Not.Roy.Astron.Soc.}\ }%
  \textbf{\bibinfo {volume} {147}},\ \bibinfo {pages} {279} (\bibinfo {year}
  {1970})%
  \bibAnnoteFile{NoStop}{Harrison}%
\bibitem{Brandenburg:2004jv}%
  \BibitemOpen
  \bibfield{author}{%
  \bibinfo {author} {\bibfnamefont{A.}~\bibnamefont{Brandenburg}}\ and\
  \bibinfo {author} {\bibfnamefont{K.}~\bibnamefont{Subramanian}},\ }%
  \bibfield{journal}{%
  \Doi{10.1016/j.physrep.2005.06.005}{\bibinfo {journal} {Phys.Rept.}}\ }%
  \textbf{\bibinfo {volume} {417}},\ \bibinfo {pages} {1} (\bibinfo {year}
  {2005}),\
  \Eprint{http://arxiv.org/abs/astro-ph/0405052}{arXiv:astro-ph/0405052
  [astro-ph]}%
  \bibAnnoteFile{NoStop}{Brandenburg:2004jv}%
\bibitem{Davis:1999bt}%
  \BibitemOpen
  \bibfield{author}{%
  \bibinfo {author} {\bibfnamefont{A.-C.}\ \bibnamefont{Davis}}, \bibinfo
  {author} {\bibfnamefont{M.}~\bibnamefont{Lilley}},\ and\ \bibinfo {author}
  {\bibfnamefont{O.}~\bibnamefont{Tornkvist}},\ }%
  \bibfield{journal}{%
  \Doi{10.1103/PhysRevD.60.021301}{\bibinfo {journal} {Phys.Rev.}}\ }%
  \textbf{\bibinfo {volume} {D60}},\ \bibinfo {pages} {021301} (\bibinfo {year}
  {1999}),\
  \Eprint{http://arxiv.org/abs/astro-ph/9904022}{arXiv:astro-ph/9904022
  [astro-ph]}%
  \bibAnnoteFile{NoStop}{Davis:1999bt}%
\bibitem{Schleicher:2010ph}%
  \BibitemOpen
  \bibfield{author}{%
  \bibinfo {author} {\bibfnamefont{D.~R.~G.}\ \bibnamefont{{Schleicher}}},
  \bibinfo {author} {\bibfnamefont{R.}~\bibnamefont{{Banerjee}}}, \bibinfo
  {author} {\bibfnamefont{S.}~\bibnamefont{{Sur}}}, \bibinfo {author}
  {\bibfnamefont{T.~G.}\ \bibnamefont{{Arshakian}}}, \bibinfo {author}
  {\bibfnamefont{R.~S.}\ \bibnamefont{{Klessen}}}, \bibinfo {author}
  {\bibfnamefont{R.}~\bibnamefont{{Beck}}},\ and\ \bibinfo {author}
  {\bibfnamefont{M.}~\bibnamefont{{Spaans}}},\ }%
  \bibfield{journal}{%
  \Doi{10.1051/0004-6361/201015184}{\bibinfo {journal} {A\&A}}\ }%
  \textbf{\bibinfo {volume} {522}},\ \bibinfo {eid} {A115} (\bibinfo {month}
  {Nov.}\ \bibinfo {year} {2010}),\
  \Eprint{http://arxiv.org/abs/1003.1135}{arXiv:1003.1135 [astro-ph.CO]}%
  \bibAnnoteFile{NoStop}{Schleicher:2010ph}%
\bibitem{Schober:2013aoa}%
  \BibitemOpen
  \bibfield{author}{%
  \bibinfo {author} {\bibfnamefont{J.}~\bibnamefont{{Schober}}}, \bibinfo
  {author} {\bibfnamefont{D.~R.~G.}\ \bibnamefont{{Schleicher}}},\ and\
  \bibinfo {author} {\bibfnamefont{R.~S.}\ \bibnamefont{{Klessen}}},\ }%
  \bibfield{journal}{%
  \Doi{10.1051/0004-6361/201322185}{\bibinfo {journal} {A\&A}}\ }%
  \textbf{\bibinfo {volume} {560}},\ \bibinfo {eid} {A87} (\bibinfo {month}
  {Dec.}\ \bibinfo {year} {2013}),\
  \Eprint{http://arxiv.org/abs/1310.0853}{arXiv:1310.0853 [astro-ph.GA]}%
  \bibAnnoteFile{NoStop}{Schober:2013aoa}%
\bibitem{BFields_Beck}%
  \BibitemOpen
  \bibfield{author}{%
  \bibinfo {author} {\bibfnamefont{R.}~\bibnamefont{Beck}}, \bibinfo {author}
  {\bibfnamefont{A.}~\bibnamefont{Brandenburg}}, \bibinfo {author}
  {\bibfnamefont{D.}~\bibnamefont{Moss}}, \bibinfo {author}
  {\bibfnamefont{A.}~\bibnamefont{Shukurov}},\ and\ \bibinfo {author}
  {\bibfnamefont{D.}~\bibnamefont{Sokoloff}},\ }%
  \bibfield{journal}{%
  \Doi{10.1146/annurev.astro.34.1.155}{\bibinfo {journal} {ARA\&A}}\ }%
  \textbf{\bibinfo {volume} {34}},\ \bibinfo {pages} {155} (\bibinfo {year}
  {1996})%
  \bibAnnoteFile{NoStop}{BFields_Beck}%
\bibitem{Balsara2001}%
  \BibitemOpen
  \bibfield{author}{%
  \bibinfo {author} {\bibfnamefont{D.}~\bibnamefont{{Balsara}}}, \bibinfo
  {author} {\bibfnamefont{R.~A.}\ \bibnamefont{{Benjamin}}},\ and\ \bibinfo
  {author} {\bibfnamefont{D.~P.}\ \bibnamefont{{Cox}}},\ }%
  \bibfield{journal}{%
  \Doi{10.1086/323967}{\bibinfo {journal} {Astrophys.J.}}\ }%
  \textbf{\bibinfo {volume} {563}},\ \bibinfo {pages} {800} (\bibinfo {month}
  {Dec.}\ \bibinfo {year} {2001}),\
  \Eprint{http://arxiv.org/abs/astro-ph/0107345}{astro-ph/0107345}%
  \bibAnnoteFile{NoStop}{Balsara2001}%
\bibitem{Sur2012}%
  \BibitemOpen
  \bibfield{author}{%
  \bibinfo {author} {\bibfnamefont{S.}~\bibnamefont{{Sur}}}, \bibinfo {author}
  {\bibfnamefont{C.}~\bibnamefont{{Federrath}}}, \bibinfo {author}
  {\bibfnamefont{D.~R.~G.}\ \bibnamefont{{Schleicher}}}, \bibinfo {author}
  {\bibfnamefont{R.}~\bibnamefont{{Banerjee}}},\ and\ \bibinfo {author}
  {\bibfnamefont{R.~S.}\ \bibnamefont{{Klessen}}},\ }%
  \bibfield{journal}{%
  \Doi{10.1111/j.1365-2966.2012.21100.x}{\bibinfo {journal} {MNRAS}}\ }%
  \textbf{\bibinfo {volume} {423}},\ \bibinfo {pages} {3148} (\bibinfo {month}
  {Jul.}\ \bibinfo {year} {2012}),\
  \Eprint{http://arxiv.org/abs/1202.3206}{arXiv:1202.3206 [astro-ph.SR]}%
  \bibAnnoteFile{NoStop}{Sur2012}%
\bibitem{Beck2013}%
  \BibitemOpen
  \bibfield{author}{%
  \bibinfo {author} {\bibfnamefont{A.~M.}\ \bibnamefont{{Beck}}}, \bibinfo
  {author} {\bibfnamefont{K.}~\bibnamefont{{Dolag}}}, \bibinfo {author}
  {\bibfnamefont{H.}~\bibnamefont{{Lesch}}},\ and\ \bibinfo {author}
  {\bibfnamefont{P.~P.}\ \bibnamefont{{Kronberg}}},\ }%
  \bibfield{journal}{%
  \Doi{10.1093/mnras/stt1549}{\bibinfo {journal} {MNRAS}}\ }%
  \textbf{\bibinfo {volume} {435}},\ \bibinfo {pages} {3575} (\bibinfo {month}
  {Nov.}\ \bibinfo {year} {2013}),\
  \Eprint{http://arxiv.org/abs/1308.3440}{arXiv:1308.3440 [astro-ph.GA]}%
  \bibAnnoteFile{NoStop}{Beck2013}%
\bibitem{Seifried2014}%
  \BibitemOpen
  \bibfield{author}{%
  \bibinfo {author} {\bibfnamefont{D.}~\bibnamefont{{Seifried}}}, \bibinfo
  {author} {\bibfnamefont{R.}~\bibnamefont{{Banerjee}}},\ and\ \bibinfo
  {author} {\bibfnamefont{D.}~\bibnamefont{{Schleicher}}},\ }%
  \bibfield{journal}{%
  \Doi{10.1093/mnras/stu294}{\bibinfo {journal} {MNRAS}}\ }%
  \textbf{\bibinfo {volume} {440}},\ \bibinfo {pages} {24} (\bibinfo {month}
  {Mar.}\ \bibinfo {year} {2014}),\
  \Eprint{http://arxiv.org/abs/1311.4991}{arXiv:1311.4991 [astro-ph.GA]}%
  \bibAnnoteFile{NoStop}{Seifried2014}%
\bibitem{Schleicher2013}%
  \BibitemOpen
  \bibfield{author}{%
  \bibinfo {author} {\bibfnamefont{D.~R.~G.}\ \bibnamefont{{Schleicher}}}\ and\
  \bibinfo {author} {\bibfnamefont{R.}~\bibnamefont{{Beck}}},\ }%
  \bibfield{journal}{%
  \Doi{10.1051/0004-6361/201321707}{\bibinfo {journal} {A\&A}}\ }%
  \textbf{\bibinfo {volume} {556}},\ \bibinfo {eid} {A142} (\bibinfo {month}
  {Aug.}\ \bibinfo {year} {2013}),\
  \Eprint{http://arxiv.org/abs/1306.6652}{arXiv:1306.6652 [astro-ph.CO]}%
  \bibAnnoteFile{NoStop}{Schleicher2013}%
\bibitem{Batchelor}%
  \BibitemOpen
  \bibfield{author}{%
  \bibinfo {author} {\bibfnamefont{G.~K.}\ \bibnamefont{{Batchelor}}},\ }%
  \bibfield{journal}{%
  \Doi{10.1098/rspa.1950.0069}{\bibinfo {journal} {Royal Society of London
  Proceedings Series A}}\ }%
  \textbf{\bibinfo {volume} {201}},\ \bibinfo {pages} {405} (\bibinfo {month}
  {Apr.}\ \bibinfo {year} {1950})%
  \bibAnnoteFile{NoStop}{Batchelor}%
\bibitem{Kazantsev}%
  \BibitemOpen
  \bibfield{author}{%
  \bibinfo {author} {\bibfnamefont{A.}~\bibnamefont{Kazantsev}},\ }%
  \bibfield{journal}{%
  \bibinfo {journal} {JHEP}\ }%
  \textbf{\bibinfo {volume} {53}},\ \bibinfo {pages} {1806} (\bibinfo {year}
  {1967})%
  \bibAnnoteFile{NoStop}{Kazantsev}%
\bibitem{Book:Kolb_Turner}%
  \BibitemOpen
  \bibfield{author}{%
  \bibinfo {author} {\bibfnamefont{E.~W.}\ \bibnamefont{Kolb}}\ and\ \bibinfo
  {author} {\bibfnamefont{M.~S.}\ \bibnamefont{Turner}},\ }%
  \emph{\bibinfo {title} {{The Early universe}}}\ (\bibinfo {publisher}
  {Addison-Wesley},\ \bibinfo {year} {1990})%
  \bibAnnoteFile{NoStop}{Book:Kolb_Turner}%
\bibitem{Banerjee:2004df}%
  \BibitemOpen
  \bibfield{author}{%
  \bibinfo {author} {\bibfnamefont{R.}~\bibnamefont{Banerjee}}\ and\ \bibinfo
  {author} {\bibfnamefont{K.}~\bibnamefont{Jedamzik}},\ }%
  \bibfield{journal}{%
  \Doi{10.1103/PhysRevD.70.123003}{\bibinfo {journal} {Phys.Rev.}}\ }%
  \textbf{\bibinfo {volume} {D70}},\ \bibinfo {pages} {123003} (\bibinfo {year}
  {2004}),\
  \Eprint{http://arxiv.org/abs/astro-ph/0410032}{arXiv:astro-ph/0410032
  [astro-ph]}%
  \bibAnnoteFile{NoStop}{Banerjee:2004df}%
\bibitem{Jedamzik:1996wp}%
  \BibitemOpen
  \bibfield{author}{%
  \bibinfo {author} {\bibfnamefont{K.}~\bibnamefont{Jedamzik}}, \bibinfo
  {author} {\bibfnamefont{V.}~\bibnamefont{Katalinic}},\ and\ \bibinfo {author}
  {\bibfnamefont{A.~V.}\ \bibnamefont{Olinto}},\ }%
  \bibfield{journal}{%
  \Doi{10.1103/PhysRevD.57.3264}{\bibinfo {journal} {Phys.Rev.}}\ }%
  \textbf{\bibinfo {volume} {D57}},\ \bibinfo {pages} {3264} (\bibinfo {year}
  {1998}),\
  \Eprint{http://arxiv.org/abs/astro-ph/9606080}{arXiv:astro-ph/9606080
  [astro-ph]}%
  \bibAnnoteFile{NoStop}{Jedamzik:1996wp}%
\bibitem{Book:Landau_Lifschitz}%
  \BibitemOpen
  \bibfield{author}{%
  \bibinfo {author} {\bibnamefont{Landau}}\ and\ \bibinfo {author}
  {\bibnamefont{Lifschitz}},\ }%
  \emph{\bibinfo {title} {{Vol. 6. Fluid mechanics}}}\ (\bibinfo {publisher}
  {Pergamon Press},\ \bibinfo {year} {1987})%
  \bibAnnoteFile{NoStop}{Book:Landau_Lifschitz}%
\bibitem{Book:Frisch}%
  \BibitemOpen
  \bibfield{author}{%
  \bibinfo {author} {\bibfnamefont{U.}~\bibnamefont{Frisch}},\ }%
  \emph{\bibinfo {title} {{Turbulence, the legacy of A.N. Kolmogorov}}}\
  (\bibinfo {publisher} {Cambridge University Press},\ \bibinfo {year} {1995})%
  \bibAnnoteFile{NoStop}{Book:Frisch}%
\bibitem{Tsuge:1974}%
  \BibitemOpen
  \bibfield{author}{%
  \bibinfo {author} {\bibfnamefont{S.}~\bibnamefont{{Tsug{\'e}}}},\ }%
  \bibfield{journal}{%
  \Doi{10.1063/1.1694592}{\bibinfo {journal} {Physics of Fluids}}\ }%
  \textbf{\bibinfo {volume} {17}},\ \bibinfo {pages} {22} (\bibinfo {month}
  {Jan.}\ \bibinfo {year} {1974})%
  \bibAnnoteFile{NoStop}{Tsuge:1974}%
\bibitem{Federrath:2011qz}%
  \BibitemOpen
  \bibfield{author}{%
  \bibinfo {author} {\bibfnamefont{C.}~\bibnamefont{Federrath}}, \bibinfo
  {author} {\bibfnamefont{G.}~\bibnamefont{Chabrier}}, \bibinfo {author}
  {\bibfnamefont{J.}~\bibnamefont{Schober}}, \bibinfo {author}
  {\bibfnamefont{R.}~\bibnamefont{Banerjee}}, \bibinfo {author}
  {\bibfnamefont{R.~S.}\ \bibnamefont{Klessen}}, \emph{et~al.},\ }%
  \bibfield{journal}{%
  \Doi{10.1103/PhysRevLett.107.114504}{\bibinfo {journal} {Phys.Rev.Lett.}}\ }%
  \textbf{\bibinfo {volume} {107}},\ \bibinfo {pages} {114504} (\bibinfo {year}
  {2011}),\ \Eprint{http://arxiv.org/abs/1109.1760}{arXiv:1109.1760
  [physics.flu-dyn]}%
  \bibAnnoteFile{NoStop}{Federrath:2011qz}%
\bibitem{SSD-Re_Pm}%
  \BibitemOpen
  \bibfield{author}{%
  \bibinfo {author} {\bibfnamefont{J.}~\bibnamefont{{Schober}}}, \bibinfo
  {author} {\bibfnamefont{D.}~\bibnamefont{{Schleicher}}}, \bibinfo {author}
  {\bibfnamefont{C.}~\bibnamefont{{Federrath}}}, \bibinfo {author}
  {\bibfnamefont{R.}~\bibnamefont{{Klessen}}},\ and\ \bibinfo {author}
  {\bibfnamefont{R.}~\bibnamefont{{Banerjee}}},\ }%
  \bibfield{journal}{%
  \Doi{10.1103/PhysRevE.85.026303}{\bibinfo {journal} {Phys.Rev. E}}\ }%
  \textbf{\bibinfo {volume} {85}},\ \bibinfo {eid} {026303} (\bibinfo {month}
  {Feb.}\ \bibinfo {year} {2012}),\
  \Eprint{http://arxiv.org/abs/1109.4571}{arXiv:1109.4571 [astro-ph.CO]}%
  \bibAnnoteFile{NoStop}{SSD-Re_Pm}%
\bibitem{Schober2012c}%
  \BibitemOpen
  \bibfield{author}{%
  \bibinfo {author} {\bibfnamefont{J.}~\bibnamefont{{Schober}}}, \bibinfo
  {author} {\bibfnamefont{D.}~\bibnamefont{{Schleicher}}}, \bibinfo {author}
  {\bibfnamefont{S.}~\bibnamefont{{Bovino}}},\ and\ \bibinfo {author}
  {\bibfnamefont{R.~S.}\ \bibnamefont{{Klessen}}},\ }%
  \bibfield{journal}{%
  \Doi{10.1103/PhysRevE.86.066412}{\bibinfo {journal} {\pre}}\ }%
  \textbf{\bibinfo {volume} {86}},\ \bibinfo {eid} {066412} (\bibinfo {month}
  {Dec.}\ \bibinfo {year} {2012}),\
  \Eprint{http://arxiv.org/abs/1212.5979}{arXiv:1212.5979 [astro-ph.CO]}%
  \bibAnnoteFile{NoStop}{Schober2012c}%
\bibitem{Federrathnum}%
  \BibitemOpen
  \bibfield{author}{%
  \bibinfo {author} {\bibfnamefont{C.}~\bibnamefont{{Federrath}}}, \bibinfo
  {author} {\bibfnamefont{S.}~\bibnamefont{{Sur}}}, \bibinfo {author}
  {\bibfnamefont{D.~R.~G.}\ \bibnamefont{{Schleicher}}}, \bibinfo {author}
  {\bibfnamefont{R.}~\bibnamefont{{Banerjee}}},\ and\ \bibinfo {author}
  {\bibfnamefont{R.~S.}\ \bibnamefont{{Klessen}}},\ }%
  \bibfield{journal}{%
  \Doi{10.1088/0004-637X/731/1/62}{\bibinfo {journal} {{Astrophys.J.}}}\ }%
  \textbf{\bibinfo {volume} {731}},\ \bibinfo {eid} {62} (\bibinfo {month}
  {Apr.}\ \bibinfo {year} {2011}),\
  \Eprint{http://arxiv.org/abs/1102.0266}{arXiv:1102.0266 [astro-ph.SR]}%
  \bibAnnoteFile{NoStop}{Federrathnum}%
\bibitem{Latif:2012aq}%
  \BibitemOpen
  \bibfield{author}{%
  \bibinfo {author} {\bibfnamefont{M.}~\bibnamefont{Latif}}, \bibinfo {author}
  {\bibfnamefont{D.}~\bibnamefont{Schleicher}}, \bibinfo {author}
  {\bibfnamefont{W.}~\bibnamefont{Schmidt}},\ and\ \bibinfo {author}
  {\bibfnamefont{J.}~\bibnamefont{Niemeyer}}}%
   (\bibinfo {year} {2012}),\
  \Eprint{http://arxiv.org/abs/1212.1619}{arXiv:1212.1619 [astro-ph.CO]}%
  \bibAnnoteFile{NoStop}{Latif:2012aq}%
\bibitem{Banerjee:2012ch}%
  \BibitemOpen
  \bibfield{author}{%
  \bibinfo {author} {\bibfnamefont{S.}~\bibnamefont{Sur}}, \bibinfo {author}
  {\bibfnamefont{D.~R.~G.}\ \bibnamefont{Schleicher}}, \bibinfo {author}
  {\bibfnamefont{R.}~\bibnamefont{Banerjee}}, \bibinfo {author}
  {\bibfnamefont{C.}~\bibnamefont{Federrath}},\ and\ \bibinfo {author}
  {\bibfnamefont{R.~S.}\ \bibnamefont{Klessen}},\ }%
  \bibfield{journal}{%
  \Doi{10.1088/2041-8205/721/2/L134}{\bibinfo {journal} {Astrophys. J.,
  Letters}}\ }%
  \textbf{\bibinfo {volume} {721}},\ \bibinfo {pages} {L134} (\bibinfo {year}
  {2010}),\ \Eprint{http://arxiv.org/abs/1008.3481}{arXiv:1008.3481
  [astro-ph.CO]}%
  \bibAnnoteFile{NoStop}{Banerjee:2012ch}%
\bibitem{Haugen:2003xp}%
  \BibitemOpen
  \bibfield{author}{%
  \bibinfo {author} {\bibfnamefont{L.}~\bibnamefont{Haugen}}, \bibinfo {author}
  {\bibfnamefont{A.}~\bibnamefont{Brandenburg}},\ and\ \bibinfo {author}
  {\bibfnamefont{W.}~\bibnamefont{Dobler}},\ }%
  \bibfield{journal}{%
  \Doi{10.1023/B:ASTR.0000045000.08395.a3}{\bibinfo {journal} {Astrophys.Space
  Sci.}}\ }%
  \textbf{\bibinfo {volume} {292}},\ \bibinfo {pages} {53} (\bibinfo {year}
  {2004}),\
  \Eprint{http://arxiv.org/abs/astro-ph/0306453}{arXiv:astro-ph/0306453
  [astro-ph]}%
  \bibAnnoteFile{NoStop}{Haugen:2003xp}%
\bibitem{Subramanian:1997pd}%
  \BibitemOpen
  \bibfield{author}{%
  \bibinfo {author} {\bibfnamefont{K.}~\bibnamefont{Subramanian}}}%
   (\bibinfo {year} {1997}),\
  \Eprint{http://arxiv.org/abs/astro-ph/9708216}{arXiv:astro-ph/9708216
  [astro-ph]}%
  \bibAnnoteFile{NoStop}{Subramanian:1997pd}%
\bibitem{Bovino-etal:2013}%
  \BibitemOpen
  \bibfield{author}{%
  \bibinfo {author} {\bibfnamefont{S.}~\bibnamefont{{Bovino}}}, \bibinfo
  {author} {\bibfnamefont{D.~R.~G.}\ \bibnamefont{{Schleicher}}},\ and\
  \bibinfo {author} {\bibfnamefont{J.}~\bibnamefont{{Schober}}},\ }%
  \bibfield{journal}{%
  \Doi{10.1088/1367-2630/15/1/013055}{\bibinfo {journal} {New Journal of
  Physics}}\ }%
  \textbf{\bibinfo {volume} {15}},\ \bibinfo {eid} {013055} (\bibinfo {month}
  {Jan.}\ \bibinfo {year} {2013}),\
  \Eprint{http://arxiv.org/abs/1212.3419}{arXiv:1212.3419 [astro-ph.CO]}%
  \bibAnnoteFile{NoStop}{Bovino-etal:2013}%
\bibitem{Schleicher:2013iz}%
  \BibitemOpen
  \bibfield{author}{%
  \bibinfo {author} {\bibfnamefont{D.~R.}\ \bibnamefont{Schleicher}}, \bibinfo
  {author} {\bibfnamefont{J.}~\bibnamefont{Schober}}, \bibinfo {author}
  {\bibfnamefont{C.}~\bibnamefont{Federrath}}, \bibinfo {author}
  {\bibfnamefont{S.}~\bibnamefont{Bovino}},\ and\ \bibinfo {author}
  {\bibfnamefont{W.}~\bibnamefont{Schmidt}},\ }%
  \bibfield{journal}{%
  \Doi{10.1088/1367-2630/15/2/023017}{\bibinfo {journal} {New J.Phys.}}\ }%
  \textbf{\bibinfo {volume} {15}},\ \bibinfo {pages} {023017} (\bibinfo {year}
  {2013}),\ \Eprint{http://arxiv.org/abs/1301.4371}{arXiv:1301.4371
  [astro-ph.CO]}%
  \bibAnnoteFile{NoStop}{Schleicher:2013iz}%
\bibitem{Gruzinov:1996gm}%
  \BibitemOpen
  \bibfield{author}{%
  \bibinfo {author} {\bibfnamefont{A.}~\bibnamefont{Gruzinov}}, \bibinfo
  {author} {\bibfnamefont{S.~C.}\ \bibnamefont{Cowley}},\ and\ \bibinfo
  {author} {\bibfnamefont{R.}~\bibnamefont{Sudan}},\ }%
  \bibfield{journal}{%
  \Doi{10.1103/PhysRevLett.77.4342}{\bibinfo {journal} {Phys.Rev.Lett.}}\ }%
  \textbf{\bibinfo {volume} {77}},\ \bibinfo {pages} {4342} (\bibinfo {year}
  {1996}),\
  \Eprint{http://arxiv.org/abs/astro-ph/9611194}{arXiv:astro-ph/9611194
  [astro-ph]}%
  \bibAnnoteFile{NoStop}{Gruzinov:1996gm}%
\bibitem{Schekochihin:2001pj}%
  \BibitemOpen
  \bibfield{author}{%
  \bibinfo {author} {\bibfnamefont{A.~A.}\ \bibnamefont{{Schekochihin}}},
  \bibinfo {author} {\bibfnamefont{S.~A.}\ \bibnamefont{{Boldyrev}}},\ and\
  \bibinfo {author} {\bibfnamefont{R.~M.}\ \bibnamefont{{Kulsrud}}},\ }%
  \bibfield{journal}{%
  \Doi{10.1086/338697}{\bibinfo {journal} {Astrophys.J.}}\ }%
  \textbf{\bibinfo {volume} {567}},\ \bibinfo {pages} {828} (\bibinfo {month}
  {Mar.}\ \bibinfo {year} {2002}),\
  \Eprint{http://arxiv.org/abs/arXiv:astro-ph/0103333}{arXiv:astro-ph/0103333}%
  \bibAnnoteFile{NoStop}{Schekochihin:2001pj}%
\bibitem{Jedamzik:1994dd}%
  \BibitemOpen
  \bibfield{author}{%
  \bibinfo {author} {\bibfnamefont{K.}~\bibnamefont{Jedamzik}}\ and\ \bibinfo
  {author} {\bibfnamefont{G.~M.}\ \bibnamefont{Fuller}},\ }%
  \bibfield{journal}{%
  \Doi{10.1086/173788}{\bibinfo {journal} {Astrophys.J.}}\ }%
  \textbf{\bibinfo {volume} {423}},\ \bibinfo {pages} {33} (\bibinfo {year}
  {1994}),\
  \Eprint{http://arxiv.org/abs/astro-ph/9312063}{arXiv:astro-ph/9312063
  [astro-ph]}%
  \bibAnnoteFile{NoStop}{Jedamzik:1994dd}%
\bibitem{Planck_param}%
  \BibitemOpen
  \bibfield{author}{%
  \bibinfo {author} {\bibfnamefont{P.}~\bibnamefont{Ade}} \emph{et~al.}
  (\bibinfo {collaboration} {Planck Collaboration})}%
   (\bibinfo {year} {2013}),\
  \Eprint{http://arxiv.org/abs/1303.5076}{arXiv:1303.5076 [astro-ph.CO]}%
  \bibAnnoteFile{NoStop}{Planck_param}%
\bibitem{Book:Lyth2009}%
  \BibitemOpen
  \bibfield{author}{%
  \bibinfo {author} {\bibfnamefont{D.~H.}\ \bibnamefont{Lyth}}\ and\ \bibinfo
  {author} {\bibfnamefont{A.~R.}\ \bibnamefont{Liddle}},\ }%
  \emph{\bibinfo {title} {{The primordial density perturbation: Cosmology,
  inflation and the origin of structure}}}\ (\bibinfo {publisher} {Cambridge
  University Press},\ \bibinfo {year} {2009})%
  \bibAnnoteFile{NoStop}{Book:Lyth2009}%
\bibitem{sasaki-cpt}%
  \BibitemOpen
  \bibfield{author}{%
  \bibinfo {author} {\bibfnamefont{H.}~\bibnamefont{Kodama}}\ and\ \bibinfo
  {author} {\bibfnamefont{M.}~\bibnamefont{Sasaki}},\ }%
  \bibfield{journal}{%
  \Doi{10.1143/PTPS.78.1}{\bibinfo {journal} {Prog.Theor.Phys.Suppl.}}\ }%
  \textbf{\bibinfo {volume} {78}},\ \bibinfo {pages} {1} (\bibinfo {year}
  {1984})%
  \bibAnnoteFile{NoStop}{sasaki-cpt}%
\bibitem{Lewis:2004kg}%
  \BibitemOpen
  \bibfield{author}{%
  \bibinfo {author} {\bibfnamefont{A.}~\bibnamefont{Lewis}},\ }%
  \bibfield{journal}{%
  \Doi{10.1103/PhysRevD.70.043518}{\bibinfo {journal} {Phys.Rev.}}\ }%
  \textbf{\bibinfo {volume} {D70}},\ \bibinfo {pages} {043518} (\bibinfo {year}
  {2004}),\
  \Eprint{http://arxiv.org/abs/astro-ph/0403583}{arXiv:astro-ph/0403583
  [astro-ph]}%
  \bibAnnoteFile{NoStop}{Lewis:2004kg}%
\bibitem{Lu:2007cj}%
  \BibitemOpen
  \bibfield{author}{%
  \bibinfo {author} {\bibfnamefont{T.~H.-C.}\ \bibnamefont{Lu}}, \bibinfo
  {author} {\bibfnamefont{K.}~\bibnamefont{Ananda}},\ and\ \bibinfo {author}
  {\bibfnamefont{C.}~\bibnamefont{Clarkson}},\ }%
  \bibfield{journal}{%
  \Doi{10.1103/PhysRevD.77.043523}{\bibinfo {journal} {Phys.Rev.}}\ }%
  \textbf{\bibinfo {volume} {D77}},\ \bibinfo {pages} {043523} (\bibinfo {year}
  {2008}),\ \Eprint{http://arxiv.org/abs/0709.1619}{arXiv:0709.1619
  [astro-ph]}%
  \bibAnnoteFile{NoStop}{Lu:2007cj}%
\bibitem{Lu:2008ju}%
  \BibitemOpen
  \bibfield{author}{%
  \bibinfo {author} {\bibfnamefont{T.~H.-C.}\ \bibnamefont{Lu}}, \bibinfo
  {author} {\bibfnamefont{K.}~\bibnamefont{Ananda}}, \bibinfo {author}
  {\bibfnamefont{C.}~\bibnamefont{Clarkson}},\ and\ \bibinfo {author}
  {\bibfnamefont{R.}~\bibnamefont{Maartens}},\ }%
  \bibfield{journal}{%
  \Doi{10.1088/1475-7516/2009/02/023}{\bibinfo {journal} {JCAP}}\ }%
  \textbf{\bibinfo {volume} {0902}},\ \bibinfo {pages} {023} (\bibinfo {year}
  {2009}),\ \Eprint{http://arxiv.org/abs/0812.1349}{arXiv:0812.1349
  [astro-ph]}%
  \bibAnnoteFile{NoStop}{Lu:2008ju}%
\bibitem{Christopherson:2010dw}%
  \BibitemOpen
  \bibfield{author}{%
  \bibinfo {author} {\bibfnamefont{A.~J.}\ \bibnamefont{Christopherson}}\ and\
  \bibinfo {author} {\bibfnamefont{K.~A.}\ \bibnamefont{Malik}},\ }%
  \bibfield{journal}{%
  \Doi{10.1088/0264-9381/28/11/114004}{\bibinfo {journal} {Class.Quant.Grav.}}\
  }%
  \textbf{\bibinfo {volume} {28}},\ \bibinfo {pages} {114004} (\bibinfo {year}
  {2011}),\ \Eprint{http://arxiv.org/abs/1010.4885}{arXiv:1010.4885 [gr-qc]}%
  \bibAnnoteFile{NoStop}{Christopherson:2010dw}%
\bibitem{Ichiki:2011ah}%
  \BibitemOpen
  \bibfield{author}{%
  \bibinfo {author} {\bibfnamefont{K.}~\bibnamefont{Ichiki}}, \bibinfo {author}
  {\bibfnamefont{K.}~\bibnamefont{Takahashi}},\ and\ \bibinfo {author}
  {\bibfnamefont{N.}~\bibnamefont{Sugiyama}},\ }%
  \bibfield{journal}{%
  \Doi{10.1103/PhysRevD.85.043009}{\bibinfo {journal} {Phys.Rev.}}\ }%
  \textbf{\bibinfo {volume} {D85}},\ \bibinfo {pages} {043009} (\bibinfo {year}
  {2012}),\ \Eprint{http://arxiv.org/abs/1112.4705}{arXiv:1112.4705
  [astro-ph.CO]}%
  \bibAnnoteFile{NoStop}{Ichiki:2011ah}%
\bibitem{Steinhardt1982}%
  \BibitemOpen
  \bibfield{author}{%
  \bibinfo {author} {\bibfnamefont{P.~J.}\ \bibnamefont{Steinhardt}},\ }%
  \bibfield{journal}{%
  \Doi{10.1103/PhysRevD.25.2074}{\bibinfo {journal} {Phys.Rev.}}\ }%
  \textbf{\bibinfo {volume} {D25}},\ \bibinfo {pages} {2074} (\bibinfo {year}
  {1982})%
  \bibAnnoteFile{NoStop}{Steinhardt1982}%
\bibitem{Kamionkowski:1993fg}%
  \BibitemOpen
  \bibfield{author}{%
  \bibinfo {author} {\bibfnamefont{M.}~\bibnamefont{Kamionkowski}}, \bibinfo
  {author} {\bibfnamefont{A.}~\bibnamefont{Kosowsky}},\ and\ \bibinfo {author}
  {\bibfnamefont{M.~S.}\ \bibnamefont{Turner}},\ }%
  \bibfield{journal}{%
  \Doi{10.1103/PhysRevD.49.2837}{\bibinfo {journal} {Phys.Rev.}}\ }%
  \textbf{\bibinfo {volume} {D49}},\ \bibinfo {pages} {2837} (\bibinfo {year}
  {1994}),\
  \Eprint{http://arxiv.org/abs/astro-ph/9310044}{arXiv:astro-ph/9310044
  [astro-ph]}%
  \bibAnnoteFile{NoStop}{Kamionkowski:1993fg}%
\bibitem{Kosowsky:2001xp}%
  \BibitemOpen
  \bibfield{author}{%
  \bibinfo {author} {\bibfnamefont{A.}~\bibnamefont{Kosowsky}}, \bibinfo
  {author} {\bibfnamefont{A.}~\bibnamefont{Mack}},\ and\ \bibinfo {author}
  {\bibfnamefont{T.}~\bibnamefont{Kahniashvili}},\ }%
  \bibfield{journal}{%
  \Doi{10.1103/PhysRevD.66.024030}{\bibinfo {journal} {Phys.Rev.}}\ }%
  \textbf{\bibinfo {volume} {D66}},\ \bibinfo {pages} {024030} (\bibinfo {year}
  {2002}),\
  \Eprint{http://arxiv.org/abs/astro-ph/0111483}{arXiv:astro-ph/0111483
  [astro-ph]}%
  \bibAnnoteFile{NoStop}{Kosowsky:2001xp}%
\bibitem{Kamionkowski1992}%
  \BibitemOpen
  \bibfield{author}{%
  \bibinfo {author} {\bibfnamefont{M.}~\bibnamefont{Kamionkowski}}\ and\
  \bibinfo {author} {\bibfnamefont{K.}~\bibnamefont{Freese}},\ }%
  \bibfield{journal}{%
  \Doi{10.1103/PhysRevLett.69.2743}{\bibinfo {journal} {Phys.Rev.Lett.}}\ }%
  \textbf{\bibinfo {volume} {69}},\ \bibinfo {pages} {2743} (\bibinfo {year}
  {1992}),\ \Eprint{http://arxiv.org/abs/hep-ph/9208202}{arXiv:hep-ph/9208202
  [hep-ph]}%
  \bibAnnoteFile{NoStop}{Kamionkowski1992}%
\bibitem{Caprini:2006jb}%
  \BibitemOpen
  \bibfield{author}{%
  \bibinfo {author} {\bibfnamefont{C.}~\bibnamefont{Caprini}}\ and\ \bibinfo
  {author} {\bibfnamefont{R.}~\bibnamefont{Durrer}},\ }%
  \bibfield{journal}{%
  \Doi{10.1103/PhysRevD.74.063521}{\bibinfo {journal} {Phys.Rev.}}\ }%
  \textbf{\bibinfo {volume} {D74}},\ \bibinfo {pages} {063521} (\bibinfo {year}
  {2006}),\
  \Eprint{http://arxiv.org/abs/astro-ph/0603476}{arXiv:astro-ph/0603476
  [astro-ph]}%
  \bibAnnoteFile{NoStop}{Caprini:2006jb}%
\bibitem{Hindmarsh:2013xza}%
  \BibitemOpen
  \bibfield{author}{%
  \bibinfo {author} {\bibfnamefont{M.}~\bibnamefont{Hindmarsh}}, \bibinfo
  {author} {\bibfnamefont{S.~J.}\ \bibnamefont{Huber}}, \bibinfo {author}
  {\bibfnamefont{K.}~\bibnamefont{Rummukainen}},\ and\ \bibinfo {author}
  {\bibfnamefont{D.~J.}\ \bibnamefont{Weir}}}%
   (\bibinfo {year} {2013}),\
  \Eprint{http://arxiv.org/abs/1304.2433}{arXiv:1304.2433 [hep-ph]}%
  \bibAnnoteFile{NoStop}{Hindmarsh:2013xza}%
\bibitem{Book:Dodelson}%
  \BibitemOpen
  \bibfield{author}{%
  \bibinfo {author} {\bibfnamefont{S.}~\bibnamefont{Dodelson}},\ }%
  \emph{\bibinfo {title} {{Modern cosmology}}}\ (\bibinfo {publisher} {Academic
  Press},\ \bibinfo {year} {2003})%
  \bibAnnoteFile{NoStop}{Book:Dodelson}%
\bibitem{Durrer:2013pga}%
  \BibitemOpen
  \bibfield{author}{%
  \bibinfo {author} {\bibfnamefont{R.}~\bibnamefont{Durrer}}\ and\ \bibinfo
  {author} {\bibfnamefont{A.}~\bibnamefont{Neronov}},\ }%
  \bibfield{journal}{%
  \Doi{10.1007/s00159-013-0062-7}{\bibinfo {journal} {Astron.Astrophys.Rev.}}\
  }%
  \textbf{\bibinfo {volume} {21}},\ \bibinfo {pages} {62} (\bibinfo {year}
  {2013}),\ \Eprint{http://arxiv.org/abs/1303.7121}{arXiv:1303.7121
  [astro-ph.CO]}%
  \bibAnnoteFile{NoStop}{Durrer:2013pga}%
\bibitem{Ahonen:1996nq}%
  \BibitemOpen
  \bibfield{author}{%
  \bibinfo {author} {\bibfnamefont{J.}~\bibnamefont{Ahonen}}\ and\ \bibinfo
  {author} {\bibfnamefont{K.}~\bibnamefont{Enqvist}},\ }%
  \bibfield{journal}{%
  \Doi{10.1016/0370-2693(96)00633-8}{\bibinfo {journal} {Phys.Lett.}}\ }%
  \textbf{\bibinfo {volume} {B382}},\ \bibinfo {pages} {40} (\bibinfo {year}
  {1996}),\ \Eprint{http://arxiv.org/abs/hep-ph/9602357}{arXiv:hep-ph/9602357
  [hep-ph]}%
  \bibAnnoteFile{NoStop}{Ahonen:1996nq}%
\bibitem{Jedamzik:2010cy}%
  \BibitemOpen
  \bibfield{author}{%
  \bibinfo {author} {\bibfnamefont{K.}~\bibnamefont{Jedamzik}}\ and\ \bibinfo
  {author} {\bibfnamefont{G.}~\bibnamefont{Sigl}},\ }%
  \bibfield{journal}{%
  \Doi{10.1103/PhysRevD.83.103005}{\bibinfo {journal} {Phys.Rev.}}\ }%
  \textbf{\bibinfo {volume} {D83}},\ \bibinfo {pages} {103005} (\bibinfo {year}
  {2011}),\ \Eprint{http://arxiv.org/abs/1012.4794}{arXiv:1012.4794
  [astro-ph.CO]}%
  \bibAnnoteFile{NoStop}{Jedamzik:2010cy}%
\bibitem{Campanelli_free-turb}%
  \BibitemOpen
  \bibfield{author}{%
  \bibinfo {author} {\bibfnamefont{L.}~\bibnamefont{Campanelli}},\ }%
  \bibfield{journal}{%
  \Doi{10.1103/PhysRevLett.98.251302}{\bibinfo {journal} {Phys.Rev.Lett.}}\ }%
  \textbf{\bibinfo {volume} {98}},\ \bibinfo {pages} {251302} (\bibinfo {year}
  {2007}),\ \Eprint{http://arxiv.org/abs/0705.2308}{arXiv:0705.2308
  [astro-ph]}%
  \bibAnnoteFile{NoStop}{Campanelli_free-turb}%
\bibitem{Brandenburg:1996fc}%
  \BibitemOpen
  \bibfield{author}{%
  \bibinfo {author} {\bibfnamefont{A.}~\bibnamefont{Brandenburg}}, \bibinfo
  {author} {\bibfnamefont{K.}~\bibnamefont{Enqvist}},\ and\ \bibinfo {author}
  {\bibfnamefont{P.}~\bibnamefont{Olesen}},\ }%
  \bibfield{journal}{%
  \Doi{10.1103/PhysRevD.54.1291}{\bibinfo {journal} {Phys.Rev.}}\ }%
  \textbf{\bibinfo {volume} {D54}},\ \bibinfo {pages} {1291} (\bibinfo {year}
  {1996}),\
  \Eprint{http://arxiv.org/abs/astro-ph/9602031}{arXiv:astro-ph/9602031
  [astro-ph]}%
  \bibAnnoteFile{NoStop}{Brandenburg:1996fc}%
\bibitem{CausalBfield}%
  \BibitemOpen
  \bibfield{author}{%
  \bibinfo {author} {\bibfnamefont{R.}~\bibnamefont{Durrer}}\ and\ \bibinfo
  {author} {\bibfnamefont{C.}~\bibnamefont{Caprini}},\ }%
  \bibfield{journal}{%
  \Doi{10.1088/1475-7516/2003/11/010}{\bibinfo {journal} {JCAP}}\ }%
  \textbf{\bibinfo {volume} {0311}},\ \bibinfo {pages} {010} (\bibinfo {year}
  {2003}),\
  \Eprint{http://arxiv.org/abs/astro-ph/0305059}{arXiv:astro-ph/0305059
  [astro-ph]}%
  \bibAnnoteFile{NoStop}{CausalBfield}%
\end{thebibliography}


%



\end{document}